\documentclass[preprint, preprintnumbers ,amsmath,amssymb]{revtex4}
\usepackage{amssymb}
\usepackage{amsmath}
\usepackage{graphicx}
\usepackage{multirow}

\begin{document}
%\preprint{APS/123-QED}

%\begin{document}

%\preprint{}

\title{Wave Function Identity: A New Symmetry for 2-electron
Systems in an Electromagnetic Field}



\author{Marlina Slamet$^{1}$ and Viraht Sahni$^{2}$}

\affiliation{$^{1}$Sacred Heart University, Fairfield, Connecticut
06825
\\
$^{2}$Brooklyn College and The Graduate School of the City
University of New York, New York, New York 10016.}


\date{\today}

\begin{abstract}
Stationary-state Schr{\"o}dinger-Pauli theory is a description of
electrons with a spin moment in an external electromagnetic field.
For 2-electron systems as described by the Schr{\"o}dinger-Pauli
theory Hamiltonian with a symmetrical binding potential, we report a
new symmetry operation of the electronic coordinates. The symmetry
operation is such that it leads to the equality of the transformed
wave function to the wave function. This equality is referred to as
the Wave Function Identity.  The symmetry operation is a two-step
process: an interchange of the spatial coordinates of the electrons
whilst keeping their spin moments unchanged, followed by an
inversion. The Identity is valid for arbitrary structure of the
binding potential, arbitrary electron interaction of the form
$w(|{\bf{r}} - {\bf{r}}'|)$, all bound electronic states, and
arbitrary dimensionality. It is proved that the exact wave functions
satisfy the Identity. On application of the permutation operation
for fermions to the identity, it is shown that the parity of the
singlet states is even and that of triplet states odd.  As a
consequence, it follows that at electron-electron coalescence, the
singlet state wave functions satisfy the cusp coalescence
constraint, and triplet state wave functions the node coalescence
condition. Further, we show that the parity of the singlet state
wave functions about all points of electron-electron coalescence is
even, and that of the triplet state wave functions odd.  The Wave
Function Identity and the properties on parity, together with the
Pauli principle, are then elucidated by application to the
$2$-dimensional $2$-electron `artificial atoms' or semiconductor
quantum dots in a magnetic field in their first excited singlet
$2^{1}S$ and triplet $2^{3} S$ states. The Wave Function Identity
and subsequent conclusions on parity are equally valid for the
special cases in which the 2-electron bound system, in both the
presence and absence of a magnetic field, are described by the
corresponding Schr{\"o}dinger theory for spinless electrons.
\end{abstract}

%\keywords{wave function functional, constraint search, variational
%principle}

\maketitle


% Section I

\section{Introduction}

This paper is concerned with a new symmetry operation of two
interacting fermions in a static electromagnetic field.  The
symmetry operation then leads to further insights into the physics
of the system.  Symmetries, and the conservation laws that result
therefrom, are fundamental to all branches of physics: high energy,
condensed matter, crystallography, atomic and molecular, and of
artificial matter.  Symmetry operations leave the physical system
invariant, and the use of symmetry facilitates the solution of
problems associated with the corresponding Hamiltonian \cite{1}. The
study of the properties of two interacting particles in diverse
physical systems has also contributed significantly to our
understanding of the underlying physics.  For interacting fermionic
systems \cite{2}, many-body theoretical methods are employed to
obtain the properties of two-particle systems, such as bound states,
scattering amplitudes, etc. by solution of the Bethe-Salpeter
equation \cite{3}.  The understanding of the net attractive
interaction near the Fermi surface at low temperatures as mitigated
by the lattice motion -- the singlet state Cooper pairs \cite{4} --
is foundational to the BCS theory of superconductivity \cite{5}.  It
is the antisymmetric wave function made up of bound Cooper pairs
that constitutes the BCS ground state \cite{5}.  Other natural bound
two-electron systems studied are the negative ion of atomic
Hydrogen, the Helium atom and its isoelectronic series
\cite{6,7,8,9,10}, and the Hydrogen molecule \cite{11,12}.
Additionally, as a result of advances in semiconductor technology,
there now exist `artificial atoms' or quantum dots, and `artificial
molecules' made up of such quantum dots.  A quantum dot differs from
a natural atom in that the electronic binding potential is harmonic
\cite{13,14,15,16,17,18}.  The natural and `artificial' two-electron
systems have also recently been studied via the `Quantal Newtonian'
first and second laws \cite{19,20,21} for the individual electron,
and by Quantal Density Functional Theory \cite{22,23}, a local
effective potential theory based on these laws.

In the present work we consider a two-electron system with
interaction of the \emph{general} form $w(|{\bf{r}} - {\bf{r}}'|)$
in an \emph{arbitrary} binding electrostatic field
$\boldsymbol{\cal{E}} ({\bf {r}}) = - \boldsymbol{\nabla} v ({\bf
{r}})/e$, where $v({\bf{r}})$ is a \emph{symmetrical} scalar
potential, and a magnetostatic field $\boldsymbol{\cal{B}} ({\bf
{r}}) = \boldsymbol{\nabla} \times {\bf{A}} ({\bf {r}})$, with
${\bf{A}} ({\bf {r}})$ the vector potential.  The system considered
is described within the context of stationary-state
Schr{\"o}dinger-Pauli theory \cite{24,25} which goes beyond
Schr{\"o}dinger theory \cite{24} in that the electron spin moment is
explicitly accounted for in the Hamiltonian. As such, the
interaction of the magnetic field with both the orbital and spin
angular momentum are considered.

We report the discovery of a new symmetry operation $O_{sym}$ about
the center of symmetry of $2$-electron systems as described by the
Schr{\"o}dinger-Pauli theory equation with \emph{arbitrary} and
\emph{even} binding potential $v ({\bf{r}})$. The symmetry operation
is such that the transformed wave function is equal to the wave
function.  The equality of the wave function to the transformed wave
function is referred to as the Wave Function Identity.  We prove
that the exact wave function of $2$-electron systems defined by the
Schr{\"o}dinger-Pauli equation satisfies this identity.  The
application of the permutation operation $P$ for fermions to the
transformed wave function then proves that the parity of all singlet
states is \emph{even}, and that of triplet states is \emph{odd}.
Thus, the product of the permutation $P$ and symmetry $O_{sym}$
operations is the inversion or parity operation $\Pi$. This, in
turn, leads to the conclusion that at electron-electron coalescence,
the singlet states satisfy the \emph{cusp} coalescence constraint,
whereas triplet states satisfy the \emph{node} coalescence
condition. Finally, it is concluded that the parity \emph{about each
point of electron-electron coalescence} in configuration space is
\emph{even} for singlet states and \emph{odd} for triplet states.

The Schr{\"o}dinger-Pauli theory eigenvalue equation for the bound
$2$-electron systems in a magnetic field ${\boldsymbol{\cal{B}}}
({\bf{r}}) = {\boldsymbol{\nabla}} \times {\bf{A}} ({\bf{r}})$ is
% Equation 1
\begin{equation}
\hat{H} \Psi ({\bf{X}}) = E \Psi ({\bf{X}}),
\end{equation}
with $\{ \Psi ({\bf{X}}), E \}$ the eigenfunctions and eigenvalues;
${\bf{X}} = {\bf{x}}_{1}, \ldots , {\bf{x}}_{N}$; ${\bf{x}} =
{\bf{r}} \sigma$; ${\bf{r}}$ and $\sigma$ the spatial and spin
coordinates. The Hamiltonian $\hat{H}$ for spin $\frac{1}{2}$
particles is comprised of the sum of the Feynman \cite{26} kinetic
$\hat{T}_{F}$, the \emph{general} electron-interaction potential
$\hat{W}$, and electrostatic binding potential $\hat{V}$ operators.
In atomic units (charge of electron $-e$; $e=\hbar=m=1$), and with
all summations $k=1$ to $2$,
% Equation 2
\begin{equation}
\hat{H} = \hat{T}_{F} + \hat{W} + \hat{V},
\end{equation}
where
% Equation 3 thru 6
\begin{eqnarray}
\hat{T}_{F} &=& \frac{1}{2} \sum_{k} \big({\boldsymbol{\sigma}}
\cdot \hat{\bf{p}}_{k, \mathrm{phys}} \big)
\big({\boldsymbol{\sigma}}
\cdot \hat{\bf{p}}_{k, \mathrm{phys}} \big) \\
&=& \frac{1}{2} \sum_{k} \big( \hat{\bf{p}}_{k} + \frac{1}{c}
{\bf{A}} ({\bf{r}}_{k}) \big)^{2} + \frac{1}{c} \sum_{k}
{\boldsymbol{\cal{B}}} ({\bf{r}}_{k}) \cdot {\bf{s}}_{k}, \\
\hat{W} &=& \frac{1}{2} \sideset{}{'}\sum_{k,\ell}
w (|{\bf{r}}_{k} - {\bf{r}}_{\ell} |) , \\
\hat{V} &=& \sum_{k} v ({\bf{r}}_{k}).
\end{eqnarray}
Here the physical momentum operator $\hat{\bf{p}}_\mathrm{phys} =
\big(\hat{\bf{p}} + \frac{1}{c} {\bf{A}} ({\bf{r}}) \big)$, with
$\hat{\bf{p}} = - i {\boldsymbol{\nabla}}$ the canonical momentum
operator; ${\boldsymbol{\sigma}}$ is the Pauli spin matrix,
${\bf{s}} = \frac{1}{2} {\boldsymbol{\sigma}}$, and ${\bf{s}}$ the
electron spin angular momentum vector operator.  The general
electron-interaction function $w(|{\bf{r}} - {\bf{r}}'|)$  could be
Coulombic, harmonic, screened-Coulomb, etc.  The binding scalar
electrostatic potential is $v ({\bf{r}})$.  For natural atoms and
molecules this potential is Coulombic, whereas for `artificial
atoms' and `artificial molecules', it is harmonic.  The use of the
Feynman kinetic energy operator leads to the correct gyromagnetic
ratio of $g=2$.

The eigenfunctions $\Psi ({\bf{X}})$ are of the form
% Equation 7
\begin{equation}
\Psi ({\bf{x}}_{1}, {\bf{x}}_{2}) \equiv \Psi ({\bf{r}}_{1}
\sigma_{1}, {\bf{r}}_{2} \sigma_{2}) = \psi ({\bf{r}}_{1},
{\bf{r}}_{2}) \chi (\sigma_{1}, \sigma_{2}),
\end{equation}
where $\psi ({\bf{r}}_{1}, {\bf{r}}_{2}), \chi (\sigma_{1},
\sigma_{2})$ are, respectively, the spatial and spin components.
That these components are separable is a consequence of the lack of
any spin-orbit interaction term in the Hamiltonian.

A property of the wave function $\Psi ({\bf{X}})$ relevant to the
present work is the constraint on it at electron-electron
coalescence.  \emph{A priori}, it is not evident which state,
singlet or triplet, satisfies the \emph{cusp} or \emph{node}
coalescence constraint.  The answer to this may be inferred from the
structure of the wave function at coalescence. With the spin
function component suppressed, the electron-electron coalescence
constraint for the spatial part of the 2-electron wave function for
dimensions $D \geq 2$ is \cite{27,28,29,30}
% Equation 8
\begin{equation}
\psi ({\bf{r}}_{1}, {\bf{r}}_{2}) = \psi({\bf{r}}_{2}, {\bf{r}}_{2})
\bigg( 1 + \frac{1} {D-1} s \bigg) + {\bf{s}} \cdot {\bf{C}}
({\bf{r}}_{2}),
\end{equation}
where ${\bf{s}} = {\bf{r}}_{2} - {\bf{r}}_{1}$, and ${\bf{C}}
({\bf{r}}_{2})$ an unknown vector.  From this (non-differential)
form of the coalescence constraint, it is possible to understand why
it is that the singlet states satisfy a cusp condition and the
triplet states a node coalescence condition.  For the singlet state,
the two electrons have opposite spin.  Hence, there is a finite
(positive-definite) probability of the two electrons being at the
same spatial position. That is, at coalescence, $\psi ({\bf{r}}_{2},
{\bf{r}}_{2})$ on the right hand side of Eq. (8) is \emph{finite}.
On the other hand, for the triplet state, the electron spins are
parallel. As a consequence of the Pauli principle, the probability
of two electrons of parallel spin being at the same physical
position is zero. Thus, $\psi ({\bf{r}}_{2}, {\bf{r}}_{2}) = 0$ in
Eq. (8), and the wave function vanishes at coalescence.

A brief summary of the properties of the wave functions $\Psi
({\bf{x}}_{1}, {\bf{x}}_{2})$, the eigenfunctions of the Hamiltonian
$\hat{H}$ of Eq. (2), obtained in the present work follows:

(a)  The wave functions satisfy the property we refer to as the Wave
Function Identity. This is achieved via a symmetry operation
$O_{sym}$, represented by the operator $\hat{O}_{sym}$, which
transforms the wave function in a two-step process on the
coordinates of the electrons: an interchange of the spatial
coordinates whilst keeping the spin moments unchanged, followed by
an inversion (reflection through the origin).  The symmetry
operation is such that the transformed wave function is equivalent
to the wave function.  This is the Wave Function Identity. It is
valid for \emph{any} $2$-electron system with an \emph{even} binding
potential and \emph{arbitrary} interaction of the form $w(|{\bf{r}}
- {\bf{r}}'|)$. It is valid for \emph{arbitrary} state whether
ground or excited, \emph{i.e.} the property is the same for both
singlet and triplet states.  It is also valid for \emph{arbitrary}
dimensionality. (Note that the switching of the electrons in the
first step of the operation $O_{sym}$ is different from that of the
Pauli principle \cite{31,32,33}.)

(b)  The application of the Pauli principle, or equivalently the
permutation operation $P_{12}$ for fermions, to the Wave Function
Identity then proves that the \emph{parity} of the singlet states is
\emph{even}, and that of the triplet states is \emph{odd}.  This
shows that the product of the permutation $P_{12}$ and symmetry
$O_{sym}$ operations is equivalent to an inversion.  In operator
form $\hat{P}_{12} \hat{O}_{sym} = \hat{\Pi}$, where $\hat{\Pi}$ is
the parity operator.

(c)  That the singlet and triplet state wave functions have even and
odd parity, respectively, then confirms that at electron-electron
coalescence the singlet state wave functions satisfy the \emph{cusp}
coalescence constraint of Eq. (8), whereas the triplet state wave
functions satisfy the \emph{node} coalescence condition.

(d)  The parity of the singlet and triplet state wave functions
about the center of symmetry further shows that \emph{about all
points of electron-electron coalescence}, the parity of the singlet
state wave functions is \emph{even}, and that of the triplet state
wave functions is \emph{odd}.

(e)   It is proved that the exact wave function of the $2$-electron
system in an \emph{arbitrary} but \emph{even} binding potential $v
({\bf{r}})$ and with an \emph{arbitrary} interaction of the form
$w(|{\bf{r}} - {\bf{r}}'|)$ must satisfy the wave function identity.

The properties of the wave functions described above in parts (a) to
(d) are then elucidated by application to the first excited singlet
$2^{1} S$ and triplet $2^{3} S$ states of a $2$D $2$-electron
`artificial atom' or semiconductor quantum dot in a magnetic field.
For these states of the `artificial atom', the \emph{exact}
solutions of the corresponding Schr{\"o}dinger-Pauli equations have
been obtained in closed analytical form \cite{19,20,21}.  As such,
the wave function properties are exhibited \emph{exactly}.  (The
Schr{\"o}dinger-Pauli Hamiltonian for the quantum dot and the
expressions for the singlet and triplet state wave functions are
given in the Appendix.)

We note that the above properties are equally valid for the
eigenfunctions $\Psi ({\bf{X}})$ of bound 2-electron systems as
described by the Schr{\"o}dinger theory of \emph{spinless}
electrons.  (By spinless electrons is meant that the spin moment of
the electron does not appear in the Hamiltonian.)  The corresponding
Hamiltonians in the presence and absence of a magnetic field, which
each correspond to a special case of Eq. (2), are respectively:
% Equation 9 & 10
\begin{eqnarray}
\hat{H}_{spinless} &=& \hat{T}_{A} + \hat{W} + \hat{V}; ~~
\hat{T}_{A} = \frac{1}{2} \sum_{k} \big( \hat{\bf{p}}_{k} +
\frac{1}{c} {\bf{A}} ({\bf{r}}_{k}) \big)^{2}, \\
\hat{H}_{spinless} &=& \hat{T} + \hat{W} + \hat{V}; ~~ \hat{T} =
\frac{1}{2} \sum_{k} \hat{p}_{k}^{2}.
\end{eqnarray}

In order to facilitate by contrast the switching of the electrons in
the first step of the symmetry operation $O_{sym}$ and that of the
Pauli principle, we begin in Sect. II by a brief discussion of the
permutation operation $P_{12}$ that leads to the Pauli principle. A
pictorial description of the Pauli principle for the first excited
singlet and triplet states of a quantum dot in a magnetic field is
also provided.  Such a representation of the Pauli principle is of
interest in its own right.  In Sect. III we describe the symmetry
operation, and explain how the Wave Function Identity is arrived at.
In Sect. IV we prove that the parity of the singlet states is even
whereas that of triplet states is odd.  We then explain in Sect. V
why the parity of singlet states about all points of
electron-electron coalescence must be even and that of triplet
states odd.  In Sect. VI we prove that the exact wave function of a
$2$-electron system in an arbitrary but symmetrical (even) binding
potential, arbitrary interaction of the form $w(|{\bf{r}} -
{\bf{r}}'|)$, and arbitrary dimensionality satisfies the Wave
Function Identity. A summary of the known properties of electronic
wave functions together with the new results of the present work is
provided in Sect. VII.

\section{Permutation Operation and the Pauli Principle}

The permutation operation $P_{12}$ permutes the coordinates
${\bf{x}}_{1}, {\bf{x}}_{2}$ of the electrons $1$ and $2$.  Fig 1 is
a $2$D vector description of the switching of the electronic
coordinates.  Fig. 1(a) corresponds to the initial coordinates of
the two electrons, and Fig. 1(b) to the switched coordinates.

The corresponding permutation operator $\hat{P}_{12}$ commutes with
the Hamiltonian $\hat{H}$:
% Equation 11
\begin{equation}
[ \hat{P}_{12}, \hat{H} ] =0,
\end{equation}
so that the eigenfunctions of $\hat{H}$ with eigenvalue $E$ (Eq.
(1)) are also eigenfunctions of the operator $\hat{P}_{12}$.  The
action of the operator $\hat{P}_{12}$ on $\Psi ({\bf{x}}_{1},
{\bf{x}}_{2})$ is thus
% Equation 12
\begin{equation}
\hat{P}_{12} \Psi ({\bf{x}}_{1}, {\bf{x}}_{2}) = \Psi ({\bf{x}}_{2},
{\bf{x}}_{1}).
\end{equation}
Operating with $\hat{P}_{12}$ on Eq. (12) one obtains
% Equation 13
\begin{equation}
\hat{P}_{12} \hat{P}_{12}  \Psi ({\bf{x}}_{1}, {\bf{x}}_{2}) = \Psi
({\bf{x}}_{1}, {\bf{x}}_{2}),
\end{equation}
so that
% Equation 14
\begin{equation}
\hat{P}_{12}^{2} = \hat{I},
\end{equation}
with $\hat{I}$ the unit operator.  Therefore the eigenvalues of the
operator $\hat{P}_{12}$ are $\epsilon = \pm 1$.  The eigenfunctions
that correspond to the eigenvalue $\epsilon = - 1$ are such that
% Equation 15
\begin{equation}
\hat{P}_{12} \Psi ({\bf{x}}_{1}, {\bf{x}}_{2}) = \Psi ({\bf{x}}_{2},
{\bf{x}}_{1}) = (-1) \Psi ({\bf{x}}_{1}, {\bf{x}}_{2})
\end{equation}
are antisymmetric under the permutation $P_{12}$.  This is the
statement of the Pauli principle:
% Equation 16
\begin{equation}
\Psi \big[ e_{1} ({\bf{r}}_{1} \sigma_{1}), e_{2} ({\bf{r}}_{2}
\sigma_{2}) \big] = - \Psi \big[ e_{1} ({\bf{r}}_{2} \sigma_{2}),
e_{2} ({\bf{r}}_{1} \sigma_{1}) \big].
\end{equation}
This statement is independent of the analytical structure or
symmetry of the binding potential.  It is independent of
dimensionality.  It is valid for arbitrary state.  For wave
functions of the form of Eq. (7), the statement of the Pauli
principle is
% Equation 17
\begin{equation}
\psi  ({\bf{r}}_{1}, {\bf{r}}_{2}) \chi (\sigma_{1}, \sigma_{2}) = -
\psi ({\bf{r}}_{2}, {\bf{r}}_{1}) \chi (\sigma_{2}, \sigma_{1}).
\end{equation}
A pictorial description of the Pauli principle for the excited
singlet $2^{1} S$ and triplet $2^{3} S$ states of a quantum dot
follows:

\emph{Singlet $2^{1} S$ state}: Employing the exact analytical
expression for the excited singlet  state of the `artificial atom'
or quantum dot as given in the Appendix Eq. (A2), the wave function
$\Psi_{S} [e_{1} ({\bf{r}}_{1} {\boldsymbol{\uparrow}}), e_{2}
({\bf{r}}_{2} {\boldsymbol{\downarrow}})] (\theta_{1} = 30^{\circ},
\theta_{2} = 65^{\circ})$ is plotted in Fig. 2(a). (This corresponds
to the case of Fig. 1(a).) The plot for the wave function when the
coordinates are switched, (corresponding to Fig. 1 (b)), which is
$\Psi_{S} [e_{1} ({\bf{r}}_{2} {\boldsymbol{\downarrow}}), e_{2}
({\bf{r}}_{1} {\boldsymbol{\uparrow}})](\theta_{1} = 65^{\circ},
\theta_{2} = 30^{\circ})$ is given in Fig. 2(b). (Note the switching
of the coordinate axes labels in Fig. 2(b).) The satisfaction of the
Pauli principle of Eq. (16) is evident.

\emph{Triplet $2^{3} S$ state}: Employing the exact analytical
expression for the triplet state wave function for the `artificial
atom' given in the Appendix by Eq. (A3), the Real part of the wave
function $\Re \Psi_{T} [e_{1} ({\bf{r}}_{1}
{\boldsymbol{\uparrow}}), e_{2} ({\bf{r}}_{2}
{\boldsymbol{\uparrow}})](\theta_{1} = 30^{\circ}, \theta_{2} =
65^{\circ})$ is plotted in Fig. 3(a). The Real part of the wave
function with the switched coordinates $\Re \Psi_{T} [e_{1}
({\bf{r}}_{2} {\boldsymbol{\uparrow}}), e_{2} ({\bf{r}}_{1}
{\boldsymbol{\uparrow}})](\theta_{1} = 65^{\circ}, \theta_{2}
=30^{\circ})$ is plotted in Fig. 3(b). In Figs 3(c) and 3(d), the
Imaginary parts of the wave function $\Im \Psi_{T} [e_{1}
({\bf{r}}_{1} {\boldsymbol{\uparrow}}), e_{2} ({\bf{r}}_{2}
{\boldsymbol{\uparrow}})](\theta_{1} = 30^{\circ}, \theta_{2} =
65^{\circ})$ and $\Im \Psi_{T} [e_{1} ({\bf{r}}_{2}
{\boldsymbol{\uparrow}}), e_{2} ({\bf{r}}_{1}
{\boldsymbol{\uparrow}})](\theta_{1} = 65^{\circ}, \theta_{2}
=30^{\circ})$, respectively, are plotted. (Note the switching of the
coordinate axes labels in Figs. 3(b) and 3(d), respectively.) The
satisfaction of the Pauli principle of Eq. (16) by the Real and
Imaginary parts of the wave function is evident.

\section{New Symmetry Operation and Wave Function Identity}

With the initial coordinates of the electrons as exhibited in Fig.
1(a), the new symmetry operation $O_{sym}$ is a two-step process on
the electron coordinates as explained in Fig. 4.  In Step 1 the
spatial coordinates of the electrons are switched $({\bf{r}}_{1}
\leftrightarrow {\bf{r}}_{2})$ while keeping the spin coordinates
associated with each electron unchanged (see quadrant 1).  Step 2 is
an inversion through the center of symmetry $({\bf{r}}_{1}
\rightarrow - {\bf{r}}_{1} ~ \mbox{and} ~ {\bf{r}}_{2} \rightarrow -
{\bf{r}}_{2})$ while the spin coordinates remain unchanged (see
quadrant 3).

The operator $\hat{O}_{sym}$ corresponding to the symmetry operation
$O_{sym}$ commutes with the Hamiltonian $\hat{H}$:
% Equation 18
\begin{equation}
\big[ \hat{O}_{sym}, \hat{H} \big] = 0,
\end{equation}
so that the eigenfunctions of $\hat{O}_{sym}$ are the same as those
of the Hamiltonian $\hat{H}$.  Operating with the operator
$\hat{O}_{sym}$ on the wave function $\Psi ({\bf{x}}_{1},
{\bf{x}}_{2})$ one obtains
% Equation 19
\begin{equation}
\hat{O}_{sym} \Psi \big[ e_{1} ({\bf{r}}_{1} \sigma_{1}), e_{2}
({\bf{r}}_{2} \sigma_{2}) \big] = \Psi \big[ e_{1} (- {\bf{r}}_{2}
\sigma_{1}), e_{2} (- {\bf{r}}_{1} \sigma_{2}) \big].
\end{equation}
Operating with $\hat{O}_{sym}$ a second time leads to
% Equation 20
\begin{equation}
\hat{O}_{sym} \hat{O}_{sym} \Psi \big[ e_{1} ({\bf{r}}_{1}
\sigma_{1}), e_{2} ({\bf{r}}_{2} \sigma_{2}) \big] = \Psi \big[
e_{1} ( {\bf{r}}_{1} \sigma_{1}), e_{2} ({\bf{r}}_{2} \sigma_{2})
\big],
\end{equation}
so that
% Equation 21
\begin{equation}
\hat{O}_{sym}^{2} = \hat{I}.
\end{equation}
Thus, the eigenvalues of the operator $\hat{O}_{sym}$ are $\lambda =
\pm 1$.  The wave functions that correspond to the eigenvalue
$\lambda = + 1$ are such that
% Equation 22
\begin{equation}
\hat{O}_{sym} \Psi \big[ e_{1} ({\bf{r}}_{1} \sigma_{1}), e_{2}
({\bf{r}}_{2} \sigma_{2}) \big] = (+1) \Psi \big[ e_{1} (
{\bf{r}}_{1} \sigma_{1}), e_{2} ({\bf{r}}_{2} \sigma_{2}) \big].
\end{equation}
The equality of the symmetry transformed wave function to the wave
function is referred to as the Wave Function Identity:
% Equation 23
\begin{equation}
\Psi \big[ e_{1} ({\bf{r}}_{1} \sigma_{1}), e_{2} ({\bf{r}}_{2}
\sigma_{2}) \big] = \Psi \big[ e_{1} (- {\bf{r}}_{2} \sigma_{1}),
e_{2} (- {\bf{r}}_{1} \sigma_{2}) \big].
\end{equation}
In terms of the wave function $\Psi ({\bf{x}}_{1}, {\bf{x}}_{2})$ of
Eq. (7), the Wave Function Identity statement is
% Equation 24
\begin{equation}
\psi ({\bf{r}}_{1}, {\bf{r}}_{2}) \chi (\sigma_{1}, \sigma_{2})  =
\psi (- {\bf{r}}_{2}, - {\bf{r}}_{1}) \chi (\sigma_{1}, \sigma_{2}).
\end{equation}
The Wave Function Identity is valid only for binding potentials $v
({\bf{r}})$ that are even.  In common with the Pauli principle, it
is valid for binding potentials of arbitrary analytical structure,
and arbitrary dimensionality.  It is also valid for both singlet and
triplet states.

A pictorial representation of the Wave Function Identity for the
singlet and triplet states of a quantum dot is provided in Figs. 5
and 6.

\emph{Singlet $2^{1} S$ state}: The Wave Function Identity of Eq.
(23) is exhibited in Fig. 5.  In Figs. 5(a) and 5(b), respectively,
the wave functions $\Psi_{S} [e_{1} ({\bf{r}}_{1}
{\boldsymbol{\uparrow}}), e_{2}({\bf{r}}_{2}
{\boldsymbol{\downarrow}})] (\theta_{1} = 30^{\circ}, \theta_{2} =
65^{\circ})$ and $\Psi_{S} [e_{1} (- {\bf{r}}_{2}
{\boldsymbol{\uparrow}}), e_{2} (-{\bf{r}}_{1}
{\boldsymbol{\downarrow}})](\theta_{1} = 245^{\circ}, \theta_{2} =
210^{\circ})$ for the `artificial atom' are plotted. (Note the
change in the labels of the coordinate axes in Fig. 5(b).) The
satisfaction of the Wave Function Identity is evident.

\emph{Triplet $2^{3} S$ state}: In Figs. 6(a) and (b), respectively,
the Real part of the triplet state wave functions for the
`artificial atom' $\Re \Psi_{T} [e_{1}({\bf{r}}_{1}
{\boldsymbol{\uparrow}}), e_{2} ({\bf{r}}_{2}
{\boldsymbol{\uparrow}})](\theta_{1} = 30^{\circ}, \theta_{2} =
65^{\circ})$ and $\Re \Psi_{T} [e_{1} (-{\bf{r}}_{2}
{\boldsymbol{\uparrow}}), e_{2} (-{\bf{r}}_{1}
{\boldsymbol{\uparrow}})](\theta_{1} = 245^{\circ}, \theta_{2} =
210^{\circ})$ are plotted. In Figs. 6(c) and (d), the Imaginary part
of the wave functions $\Im \Psi_{T} [e_{1} ({\bf{r}}_{1}
{\boldsymbol{\uparrow}}), e_{2} ({\bf{r}}_{2}
{\boldsymbol{\uparrow}})](\theta_{1} = 30^{\circ}, \theta_{2} =
65^{\circ})$ and $\Im \Psi_{T} [e_{1} (-{\bf{r}}_{2}
{\boldsymbol{\uparrow}}), e_{2} (-{\bf{r}}_{1}
{\boldsymbol{\uparrow}})](\theta_{1} = 245^{\circ}, \theta_{2} =
210^{\circ})$ are plotted. (Note the change in the labels of the
coordinate axes in Figs. 6(b) and 6(d).)These figures demonstrate
the satisfaction of the Wave Function Identity for the Real and
Imaginary parts of the wave function for the triplet state.

\section{Parity of Singlet and Triplet State Wave Functions}

We next show by application of the permutation operator
$\hat{P}_{12}$ to the Wave Function Identity that the parity of the
singlet states is even, and that of the triplet states odd.  To do
so, consider the form of the wave function $\Psi ({\bf{x}}_{1},
{\bf{x}}_{2})$ to be that of Eq. (7).  We know from the previous
section that
% Equation 25
\begin{equation}
\hat{O}_{sym} \big[ \psi ({\bf{r}}_{1}, {\bf{r}}_{2}) \chi
(\sigma_{1}, \sigma_{2}) \big] = \psi (- {\bf{r}}_{2}, -
{\bf{r}}_{1}) \chi (\sigma_{1}, \sigma_{2}).
\end{equation}
On application of the permutation operator $\hat{P}_{12}$ to Eq.
(25), we have
% Equation 26
\begin{equation}
\hat{P}_{12} \hat{O}_{sym} \big[ \psi ({\bf{r}}_{1}, {\bf{r}}_{2})
\chi (\sigma_{1}, \sigma_{2}) \big] = - \psi (- {\bf{r}}_{1}, -
{\bf{r}}_{2}) \chi (\sigma_{2}, \sigma_{1}).
\end{equation}
(Note that the negative sign on the right is a consequence of the
Pauli principle.)  Let us consider the singlet and triplet states
separately.

\emph{Singlet $2^{1} S$ state}:  For the singlet state $\Psi_{S}
({\bf{x}}_{1}, {\bf{x}}_{2})$, the spin component of the wave
function $\chi_{S} (\sigma_{1}, \sigma_{2})$ is antisymmetric in an
interchange of the spin coordinates.  Hence, the spatial component
$\psi_{S} ({\bf{r}}_{1}, {\bf{r}}_{2})$ is symmetric in an
interchange of the spatial coordinates.  Employing these
constraints, we then have from Eq. (24)
% Equation 27, 28, 29 and 30
\begin{eqnarray}
\hat{P}_{12} \hat{O}_{sym} \big[ \psi_{S} ({\bf{r}}_{1},
{\bf{r}}_{2}) \chi_{S} (\sigma_{1}, \sigma_{2}) \big] &=& - \psi_{S}
(- {\bf{r}}_{1},
- {\bf{r}}_{2}) \chi_{S} (\sigma_{2}, \sigma_{1}) \\
&=& \big[- \psi_{S} (- {\bf{r}}_{1}, - {\bf{r}}_{2}) \big]
\big[- \chi_{S}  (\sigma_{1}, \sigma_{2}) \big]\\
&=& \psi_{S} (- {\bf{r}}_{1}, - {\bf{r}}_{2}) \chi_{S} (\sigma_{1},
\sigma_{2}) \\
&=& \psi_{S} (- {\bf{r}}_{2}, - {\bf{r}}_{1}) \chi_{S} (\sigma_{1},
\sigma_{2}).
\end{eqnarray}
From a comparison of the Wave Function Identity of Eq. (24) to that
of Eq. (30), together with Eq. (29), it follows that for the singlet
state
% Equation 31
\begin{equation}
\psi_{S} ({\bf{r}}_{1}, {\bf{r}}_{2}) \chi (\sigma_{1}, \sigma_{2})
= \psi_{S} (- {\bf{r}}_{1}, - {\bf{r}}_{2}) \chi_{S} (\sigma_{1},
\sigma_{2})
\end{equation}
or more generally
% Equation 32
\begin{equation}
\Psi_{S} ({\bf{x}}_{1}, {\bf{x}}_{2}) = \Psi_{S} (- {\bf{x}}_{1}, -
{\bf{x}}_{2}),
\end{equation}
where $ - {\bf{x}} = - {\bf{r}} \sigma$.  The right hand side of Eq.
(31) corresponds to an inversion about the center of symmetry, and
thus the singlet state has even parity.  In Fig. 7 we demonstrate
the even parity of the singlet state wave function for the
`artificial atom' by plotting $\Psi_{S} [e_{1} ({\bf{r}}_{1}
{\boldsymbol{\uparrow}}), e_{2} ({\bf{r}}_{2}
{\boldsymbol{\downarrow}})] (\theta_{1} = 30^{\circ}; \theta_{2} =
65^{\circ})$ and $\Psi_{S} [e_{1} (- {\bf{r}}_{1}
{\boldsymbol{\uparrow}}), e_{2} (- {\bf{r}}_{2}
{\boldsymbol{\downarrow}})] (\theta_{1} = 210^{\circ}; \theta_{2} =
245^{\circ})$.

\emph{Triplet $2^{3} S$ state}:  For the triplet state, the spin
component of the wave function $\chi_{T} (\sigma_{1}, \sigma_{2})$
is symmetric in an interchange of the spin coordinates, and
therefore the spatial component $\psi_{T} ({\bf{r}}_{1},
{\bf{r}}_{2})$ is antisymmetric in an interchange of the spatial
coordinates.  Employing these constraints, we then have from Eq.
(26)
% Equation 33, 34, 35 and 36
\begin{eqnarray}
\hat{P}_{12} \hat{O}_{sym} \big[ \psi_{T} ({\bf{r}}_{1},
{\bf{r}}_{2}) \chi_{T} (\sigma_{1}, \sigma_{2}) \big] &=& - \psi_{T}
(-
{\bf{r}}_{1}, - {\bf{r}}_{2}) \chi_{T} (\sigma_{2}, \sigma_{1}) \\
&=& - \psi_{T} (- {\bf{r}}_{1}, - {\bf{r}}_{2})
\chi_{T}  (\sigma_{1}, \sigma_{2})  \\
&=& -[ - \psi_{T} (- {\bf{r}}_{2}, - {\bf{r}}_{1})] \chi_{T}
(\sigma_{1}, \sigma_{2}) \\
&=& \psi_{T} (- {\bf{r}}_{2}, - {\bf{r}}_{1}) \chi_{T} (\sigma_{1},
\sigma_{2}).
\end{eqnarray}
From a comparison of the Wave Function Identity of Eq. (24) to that
of Eq. (36), together with Eq. (34), it follows that for the triplet
state
% Equation 37
\begin{equation}
\psi_{T} ({\bf{r}}_{1}, {\bf{r}}_{2}) \chi (\sigma_{1}, \sigma_{2})
= - \psi_{T} (- {\bf{r}}_{1}, - {\bf{r}}_{2}) \chi_{T} (\sigma_{1},
\sigma_{2})
\end{equation}
or more generally
% Equation 38
\begin{equation}
\Psi_{T} ({\bf{x}}_{1}, {\bf{x}}_{2}) = - \Psi_{T}  (- {\bf{x}}_{1},
- {\bf{x}}_{2}).
\end{equation}
Since on inversion, the wave function changes sign, the parity of
the triplet states is odd.

 To exhibit the odd parity of the triplet state wave
function, we plot in Fig. 8(a) $\Re \Psi_{T} [e_{1} ({\bf{r}}_{1}
{\boldsymbol{\uparrow}}), e_{2} ({\bf{r}}_{2}
{\boldsymbol{\uparrow}})] (\theta_{1} = 30^{\circ}, \theta_{2} =
65^{\circ})$ and $\Re \Psi_{T} [e_{1} (-{\bf{r}}_{1}
{\boldsymbol{\uparrow}}), e_{2} (-{\bf{r}}_{2}
{\boldsymbol{\uparrow}})](\theta_{1} = 210^{\circ}, \theta_{2} =
245^{\circ})$ . In Fig. 8(b) we plot $\Im \Psi_{T} [e_{1}
({\bf{r}}_{1} {\boldsymbol{\uparrow}}), e_{2} ({\bf{r}}_{2}
{\boldsymbol{\uparrow}})] (\theta_{1} = 30^{\circ}, \theta_{2} =
65^{\circ})$ and $\Im \Psi_{T} [e_{1} (-{\bf{r}}_{1}
{\boldsymbol{\uparrow}}), e_{2} (-{\bf{r}}_{2}
{\boldsymbol{\uparrow}})] (\theta_{1} = 210^{\circ}, \theta_{2} =
245^{\circ})$.

It is evident from the above that the symmetry operation $O_{sym}$
followed by a permutation $P_{12}$ is equivalent to an inversion
(see Fig. 9).  In operator form,
% Equation 39
\begin{equation}
\hat{P}_{12} \hat{O}_{sym} = \hat{\Pi},
\end{equation}
where $\hat{\Pi}$ is the parity operator.  The properties of the
parity operator are the following:  $[\hat{\Pi}, \hat{H}] = 0$;
$\hat{\Pi}^{2} = \hat{I}$; the eigenvalues of $\hat{\Pi}$ are
$\alpha = \pm 1$.

As a consequence of Eqs. (32), (38), (39), we conclude that the
eigenfunctions of the parity operator for eigenvalue $\alpha = 1$
are singlet states, and those of the eigenvalue $\alpha = -1$ are
triplet states.  Thus,
% Equation 40
\begin{equation}
\hat{P}_{12} \hat{O}_{sym} \Psi_{S} ({\bf{r}}_{1} \sigma_{1},
{\bf{r}}_{2} \sigma_{2}) = \hat{\Pi} \Psi_{S} ({\bf{r}}_{1}
\sigma_{1}, {\bf{r}}_{2} \sigma_{2}) = \Psi_{S} (- {\bf{r}}_{1}
\sigma_{1}, {- \bf{r}}_{2} \sigma_{2}),
\end{equation}
and
% Equation 41
\begin{equation}
\hat{P}_{12} \hat{O}_{sym} \Psi_{T} ({\bf{r}}_{1} \sigma_{1},
{\bf{r}}_{2} \sigma_{2}) = \hat{\Pi} \Psi_{T} ({\bf{r}}_{1}
\sigma_{1}, {\bf{r}}_{2} \sigma_{2}) = - \Psi_{T} (- {\bf{r}}_{1}
\sigma_{1}, {- \bf{r}}_{2} \sigma_{2}).
\end{equation}

Note that \emph{a priori} it is not known that the positive
eigenvalue of the parity operator $\hat{\Pi}$ corresponds to the
singlet state, whereas the negative eigenvalue that of the triplet
state.  It is only \emph{a posteriori}, that is following the proof
that the parity of singlet states is even and that of triplet states
odd, that one can associate the positive eigenvalue of the parity
operator $\hat{\Pi}$ with singlet states and the negative eigenvalue
with the triplet state.

\section{Parity about all Points of Electron-Electron Coalescence}

As explained in the Introduction, one may conclude from the
electron-electron coalescence condition of Eq.(8), that the singlet
states satisfy the cusp coalescence constraint whereas the triplet
states satisfy the node coalescence condition. In other words, the
singlet state wave functions exhibit a cusp at \emph{all} points of
electron-electron coalescence, and triplet states exhibit a node at
these points.  We have also proved that the singlet states have even
parity and the triplet states odd parity.  The parity of a wave
function is with respect to the center (origin) of the system as
defined by the symmetrical binding potential. Now the origin of the
system also constitutes a point of electron-electron coalescence.
Thus for singlet states, the cusp at the origin must be such that
the slope of the wave function as the origin is approached from the
right must be the same but the negative of the slope when approached
from the left.   The slopes have a discontinuity at the origin. From
the perspective of the coalescence of electrons, the origin is not a
special point of configuration space.  There is no reason why the
parity about all other points of electron-electron coalescence
should not also be the same.  Hence, it follows that the parity of
the singlet state wave functions \emph{at all points of
electron-electron coalescence} is even.

To show that this is the case, we plot in Fig. 10 the singlet $2^{1}
S$ state wave function of the `artificial atom' as a function of the
center of mass ${\bf{R}} = ({\bf{r}}_{1} + {\bf{r}}_{2})/2$ and
relative ${\bf{s}} = {\bf{r}}_{2} - {\bf{r}}_{1}$ coordinates. As
may be observed from the figure, the parity of the wave function
about the line ${\bf{s}} = 0$ is even.

The triplet state wave functions exhibit a node at the origin, and
have odd parity.  Hence, the functions are smooth, \emph{i.e}.
continuous and with continuous first derivatives, about the origin.
Again, the origin corresponds to just one point of electron-electron
coalescence.  Thus, one may conclude that triplet state wave
functions have odd parity about \emph{all points of
electron-electron coalescence}.  In Fig. 11 (a) and (b),  we plot
two views of the Real part of the triplet $2^{3} S$ state wave
function of the `artificial atom' as a function of the coordinates
$({\bf{s}}, {\bf{R}})$ for $\alpha = 30^{\circ}$, where $\alpha$ is
the angle of the relative coordinate. Observe that the parity of the
wave function about the line ${\bf{s}} = 0$ is odd. In Fig. 11 (c)
and (d), two views of the Imaginary part of the wave function are
plotted. Again, the parity about the line ${\bf{s}} = 0$ is odd.

Finally, implicit in the understanding that the singlet state wave
functions have even parity is that they satisfy the \emph{cusp}
electron-electron coalescence constraint.  This too is evident in
Fig. 10 for all points ${\bf{s}} = 0$.   The fact that triplet
states have odd parity then implies that these wave functions
satisfy the \emph{node} coalescence condition.  This is evident for
both the Real and Imaginary parts of the triplet state wave function
in Fig. 11 along the ${\bf{s}} = 0$ line.

\section{Proof of Satisfaction of the Wave Function Identity by the Exact Wave
Function}

In this section we prove that the exact wave function of any
$2$-electron system with symmetrical binding potential $v
({\bf{r}})$ must satisfy the Wave Function Identity.  The examples
of the previous section show that this is the case for the $2$D
`artificial atoms' with symmetric harmonic binding potential.  This
is also the case for $3$D `artificial atoms' with the magnetic field
absent.  This can be seen from the solutions of the corresponding
Schr{\"o}dinger equations which can also be obtained \emph{exactly}
in closed analytical form \cite{22,23,34,35,36,37,38,39}.  Here we
prove the result for arbitrary symmetrical potential $v ({\bf{r}})$
and arbitrary interaction of the form $w(|{\bf{r}} - {\bf{r}}'|)$.

The exact wave function $\Psi ({\bf{x}}_{1}, {\bf{x}}_{2})$ of Eq.
(7) may be written as
% Equation 42
\begin{equation}
\Psi ({\bf{x}}_{1}, {\bf{x}}_{2}) = \bigg[ \sum_{i} c_{i} \Phi_{i}
({\bf{r}}_{1}, {\bf{r}}_{2}) \bigg] \chi (\sigma_{1}, \sigma_{2}),
\end{equation}
so that the spatial part $\psi ({\bf{r}}_{1}, {\bf{r}}_{2})$ is
given by the infinite series expansion
% Equation 43
\begin{equation}
\psi ({\bf{r}}_{1}, {\bf{r}}_{2}) = \sum_{i} c_{i} \Phi_{i}
({\bf{r}}_{1}, {\bf{r}}_{2}),
\end{equation}
where the determinantal functions $\Phi_{i} ({\bf{r}}_{1},
{\bf{r}}_{2})$ form a complete orthonormal set, with the
coefficients $c_{i}$ being suitably chosen constants.  If it is
ensured that each determinant $\Phi_{i} ({\bf{r}}_{1},
{\bf{r}}_{2})$ satisfies the Wave Function Identity, the exact wave
function will satisfy the identity.

\emph{Triplet States}

Since for the triplet state, the spin component of the wave function
$\chi (\sigma_{1}, \sigma_{2})$  is symmetric in an interchange of
the spin coordinates $(\sigma_{1}, \sigma_{2})$, each spatial
component determinant $\Phi_{i} ({\bf{r}}_{1}, {\bf{r}}_{2})$ must
be antisymmetric in an interchange of the spatial coordinates
$({\bf{r}}_{1}, {\bf{r}}_{2})$. Thus, an arbitrary determinant $\Phi
({\bf{r}}_{1}, {\bf{r}}_{2})$ is given as
% Equation 44
\begin{equation}
\Phi ({\bf{r}}_{1}, {\bf{r}}_{2}) = \phi_{1} ({\bf{r}}_{1}) \phi_{2}
({\bf{r}}_{2}) - \phi_{2} ({\bf{r}}_{1}) \phi_{1} ({\bf{r}}_{2}).
\end{equation}
As the binding potential $v({\bf{r}})$ is symmetrical, it is
possible to ensure that the orbital $\phi_{1} ({\bf{r}})$ has
\emph{even} parity and the orbital $\phi_{2} ({\bf{r}})$ has
\emph{odd} parity, \emph{i.e}.
% Equation 45
\begin{equation}
\phi_{1} ({\bf{r}}) = \phi_{1} (-{\bf{r}})~;~~~~ \phi_{2} ({\bf{r}})
= - \phi_{2} (-{\bf{r}}).
\end{equation}
Then, the spatial part of the right hand side of the Wave Function
Identity of Eq. (24) (with $\psi (- {\bf{r}}_{2}, - {\bf{r}}_{1})$
replaced by $\Phi (- {\bf{r}}_{2}, - {\bf{r}}_{1})$) employing Eq.
(44) becomes
% Equation 46
\begin{equation}
\Phi (-{\bf{r}}_{2}, - {\bf{r}}_{1}) = \phi_{1} (-{\bf{r}}_{2})
\phi_{2} (-{\bf{r}}_{1}) - \phi_{2}  (-{\bf{r}}_{2}) \phi_{1} (-
{\bf{r}}_{1}).
\end{equation}
On substituting Eq. (45) on the right hand side of Eq. (46), we have
% Equation 47 thru 49
\begin{eqnarray}
\Phi (-{\bf{r}}_{2}, - {\bf{r}}_{1}) &=& \phi_{1} ({\bf{r}}_{2})
[-\phi_{2} ({\bf{r}}_{1})]  - [- \phi_{2} ({\bf{r}}_{2})] \phi_{1}
({\bf{r}}_{1}) \\
&=& \phi_{1} ({\bf{r}}_{1}) \phi_{2} ({\bf{r}}_{2})  - \phi_{2}
({\bf{r}}_{1})  \phi_{1} ({\bf{r}}_{2}) \\
&=& \Phi ({\bf{r}}_{1}, {\bf{r}}_{2}).
\end{eqnarray}
This proves that the Wave Function Identity is satisfied for each
determinant $\Phi_{i} ({\bf{r}}_{1}, {\bf{r}}_{2})$ of $\psi
({\bf{r}}_{1}, {\bf{r}}_{2})$ of Eq. (43), and thereby by the
\emph{exact} triplet state wave function $\Psi ({\bf{x}}_{1},
{\bf{x}}_{2})$ of Eq. (42).

\emph{Singlet States}

For the singlet state, the spin component of the wave function $\chi
(\sigma_{1}, \sigma_{2})$  is antisymmetric in an interchange of the
spin coordinates $(\sigma_{1}, \sigma_{2})$.  Thus, each spatial
component determinant $\Phi_{i} ({\bf{r}}_{1}, {\bf{r}}_{2})$  must
be symmetric in an interchange of the spatial coordinates
$({\bf{r}}_{1}, {\bf{r}}_{2})$. Hence, the determinant $\Phi_{i}
({\bf{r}}_{1}, {\bf{r}}_{2})$ is given as
% Equation 50
\begin{equation}
\Phi_{i} ({\bf{r}}_{1},  {\bf{r}}_{2}) = \phi_{1} ({\bf{r}}_{1})
\phi_{2} ({\bf{r}}_{2}) + \phi_{2} ({\bf{r}}_{1})  \phi_{1}
({\bf{r}}_{2}).
\end{equation}
If one now ensures that \emph{both} orbitals $\phi_{1} ({\bf{r}})$
and $\phi_{2} ({\bf{r}})$ have \emph{even} parity, then once again
following the procedure above, we have
% Equation 51
\begin{equation}
\Phi (-{\bf{r}}_{2}, - {\bf{r}}_{1}) = \Phi ({\bf{r}}_{1},
{\bf{r}}_{2}),
\end{equation}
Thus, each determinant $\Phi_{i} ({\bf{r}}_{1},  {\bf{r}}_{2})$, and
therefore the \emph{exact} singlet state wave function $\Psi
({\bf{x}}_{1}, {\bf{x}}_{2})$, too satisfies the Wave Function
Identity.

What one also learns from the above, is that in the construction of
approximate configuration-interaction type wave functions of the
form of Eq. (42), one orbital $\phi_{1} ({\bf{r}})$ must always have
even parity, whereas the other $\phi_{2} ({\bf{r}})$ must have even
parity for singlet states and odd parity for triplet states. This
will ensure not only the satisfaction of the Wave Function Identity,
but also that the approximate singlet and triplet state wave
functions will have even and odd parity, respectively, as they must.
The approximate wave functions will then also have the correct
parity about each point of electron-electron coalescence.

\section{Summary of Results}

For a bound $N$-electron system in the presence of a magnetic field
as described by the Schr{\"o}dinger-Pauli theory Hamiltonian which
explicitly accounts for the spin moment of the electron, the wave
function $\Psi ({\bf{X}})$ possesses certain properties.  Any
approximation to the wave function must then be so constrained.
These constraints are the following: (a) must be continuous, single
valued, and bounded; (b) satisfy the Pauli principle; (c) be
normalized with probability density $\geq 0$; (d) satisfy either the
cusp or node electron-electron coalescence condition; (e) satisfy
the electron-nucleus coalescence constraint for binding potentials
that are singular at the nucleus; (f) possess the appropriate number
of nodes and the correct asymptotic structure in the classically
forbidden region; (g) have the correct parity.  There are then
additional properties of the wave function in momentum space
\cite{40,41}.

For $2$-electron systems as described by the Schr{\"o}dinger-Pauli
equation with a symmetrical binding potential, we have discovered
the following additional properties and facets of the wave function:

\emph{\textbf{(i)}}  A new symmetry operation which leads to the
equality of the transformed wave function to the wave function,
referred to as the Wave Function identity, has been discovered.  In
common with the Pauli principle, the Identity is valid for both
singlet and triplet states; for arbitrary analytical structure of
the binding potential; arbitrary interaction of the form
$w(|{\bf{r}} - {\bf{r}}'|)$; and for arbitrary dimensionality.

\emph{\textbf{(ii)}}  On application of the Pauli principle to the
Wave Function Identity, it is shown that the parity of singlet
states is even, and that of triplet states is odd.  (\emph{A
priori}, the parity of these states is not evident, unless the
Schr{\"o}dinger-Pauli equation is solved in closed analytical form.
The property of parity is thus not emphasized in the literature.)

\emph{\textbf{(iii)}} As a consequence of the parity, \emph{at
electron-electron coalescence}, the singlet state wave functions
satisfy the cusp coalescence constraint, whereas the triplet state
wave functions satisfy the node coalescence condition.  (The parity
argument constitutes an independent way of arriving at this
conclusion.)

\emph{\textbf{(iv)}} Further, the parity of the wave function about
\emph{each point of electron-electron coalescence} is even for
singlet states and odd for triplet states.  (To our knowledge, this
parity is not described in the literature.  There is, however,
substantial literature on the requirement of the satisfaction of the
electron-electron coalescence constraints on approximate wave
functions \cite{42}.  Additionally, there is work on the
electron-nucleus coalescence constraint in differential form in
terms of the electron density for Coulombic external potentials, see
e.g. \cite{43}.)

\emph{\textbf{(v)}} Finally, it is proved that the \emph{exact} wave
function satisfies the Wave Function Identity.

The above properties, including the satisfaction of the Pauli
principle, are elucidated for both the singlet $2^{1} S$ and triplet
$2^{3} S$ states by application to the $2$-electron $2$D `artificial
atoms' or semiconductor quantum dots in a magnetic field.  As the
solutions of the corresponding Schr{\"o}dinger-Pauli equation are
known in closed analytical form, the description of the above
properties is \emph{exact}.

Note that the Schr{\"o}dinger theory of \emph{spinless} electrons,
both in the presence and absence of a magnetic field, constitute
special cases of Schr{\"o}dinger-Pauli theory. As such all the above
properties are equally valid for these separate descriptions of the
system.

In conclusion, we now have an additional constraint -- the Wave
Function Identity -- that must be satisfied by any approximate
$2$-electron wave function.  This will ensure that the approximate
wave function has the correct parity, and the correct parity about
each point of electron-electron coalescence.

We are presently investigating whether the Wave Function Identity is
valid in general for solutions of the Schr{\"o}dinger-Pauli equation
for $N \geq 3$.

\textbf{Acknowledgement}: The authors thank Prof Xiaoyin Pan for his
critique of the paper.

\clearpage

\appendix

\section{The `Artificial Atom' or Quantum Dot in a Magnetic Field}

The physical system employed to exhibit the various properties of
the wave function described in the text is the $2$D $2$-electron
`artificial atom' or semiconductor quantum dot in a magnetic field
\cite{44,45,46,47,48,49,50}.  The motion of the electrons is
confined to two dimensions in a quantum well within a thin layer of
semiconductor such as GaAs which is sandwiched between two much
thicker layers of another semiconductor AlGaAs.  The lateral
confinement is achieved by placing an electrostatic gate on this
system.  The electrons are further constrained by application of a
magnetic field perpendicular to the plane of motion.  For the
`artificial atom' the free electron mass $m$ is replaced by the
semiconductor band effective mass $m^{\star}$, and the
electron-interaction modified by the dielectric constant $\epsilon$.

In contrast to natural atoms, the electrons in the `artificial atom'
are bound by a harmonic potential $v ({\bf{r}}) = \frac{1}{2}
m^{\star} \omega^{2}_{0} r^{2}$, with $\omega_{0}$ the harmonic
frequency.  The Schr{\"o}dinger-Pauli equation is then
% Equation A1
\begin{eqnarray}
\bigg[ \frac{1}{2 m^{\star}} \sum^{2}_{k=1} \big(\hat{\bf{p}}_{k}
&+& \frac{e} {c} {\bf{A}} ({\bf{r}}_{k}) \big)^{2} + g^{\star}
\mu_{B} \sum^{2}_{k=1} {\boldsymbol{\cal{B}}} ({\bf{r}}_{k}) \cdot
{\bf{s}}_{k} + \frac{1}{\epsilon} \frac{e^{2}} {|{\bf{r}}_{1} -
{\bf{r}}_{2}|}  \nonumber \\
&+& \frac{1}{2} m^{\star}\omega^{2}_{0} \sum^{2}_{k=1} r^{2}_{k}
\bigg] \Psi ({\bf{x}}_{1}, {\bf{x}}_{2}) = E \Psi ({\bf{x}}_{1},
{\bf{x}}_{2}),
\end{eqnarray}
where $\mu_{B} = e \hbar/2m$ is the Bohr magneton; $g^{\star}$ the
gyromagnetic ratio; ${\bf{x}} = {\bf{r}} \sigma$; ${\bf{r}} = (r
\theta)$; ${\bf{r}} \sigma$ the spatial and spin coordinates.

In the symmetric gauge ${\bf{A}} ({\bf{r}}) = \frac{1}{2}
{\boldsymbol{\cal{B}}} ({\bf{r}}) \times {\bf{r}}$ with the magnetic
field in the $z$-direction ${\boldsymbol{\cal{B}}} ({\bf{r}}) =
{\cal{B}} {\bf{i}}_{z}$, the Schr{\"o}dinger-Pauli equation can be
solved \cite{19,20,21} in closed analytical form by the method of
Taut \cite{44,45,46} for the first excited singlet $2^{1} S$ and
triplet $2^{3} S$ states for a denumerably infinite set of
frequencies $\omega_{0}$ and $\omega_{L}$ such that the effective
force constant $k_\mathrm{eff} = \omega^{2}_{0} + \omega^{2}_{L}$,
where $\omega_{L} = {\cal{B}}/2c$ is the Larmor frequency. Effective
atomic units are employed: $e^{2}/\epsilon =\hbar = m^{\star} =
c=1$.  The effective Bohr radius is $a^{\star}_{0} = a_{0}
(m/m^{\star})$, the effective energy unit is $(a.u.)^{\star} =
(a.u.) (m^{\star}/mc^{2})$.  The wave functions are of the form
$\Psi ({\bf{x}}_{1}, {\bf{x}}_{2}) = \psi ({\bf{r}}_{1},
{\bf{r}}_{2}) \chi (\sigma_{1}, \sigma_{2})$ with $\psi
({\bf{r}}_{1}, {\bf{r}}_{2})$ the spatial and $\chi (\sigma_{1},
\sigma_{2})$ the spin components.  The expressions for the spatial
components $\psi ({\bf{r}}_{1}, {\bf{r}}_{2})$ of the singlet and
triplet states and their respective energies $E$ are given below.

\newpage

\emph{Singlet $2^{1} S$ State}
% Equation A2
\begin{eqnarray}
\psi ({\bf{r}}_{1}, {\bf{r}}_{2}) &=& Ne^{- {\sqrt{k_\mathrm{eff}}}
(r^{2}_{1} + r^{2}_{2})/{2}} [1 + |{\bf{r}}_{1} - {\bf{r}}_{2}| +
c_{2} |{\bf{r}}_{1} - {\bf{r}}_{2}|^{2} + c_{3} |{\bf{r}}_{1} -
{\bf{r}}_{2}|^{3}] \nonumber \\
N &=& 0.108563 ~~~ \mbox{(Normalization constant)}  \nonumber \\
k_\mathrm{eff} &=& 0.471716 ~~~\mbox{(Effective force constant)}
\nonumber \\
c_{2} &=& - 0.265111; ~~c_{3} = -0.182082 ~~~\mbox{(Coefficients
of expansion)} \nonumber \\
E &=& 3.434066 ~(a.u.)^{\star} ~~~ \mbox{(Energy)}.
\end{eqnarray}
Observe that since the spin component $\chi(\sigma_{1}, \sigma_{2})$
is antisymmetric in an interchange of the spin coordinates
$\sigma_{1}$ and $\sigma_{2}$, the spatial component $\psi
({\bf{r}}_{1}, {\bf{r}}_{2})$ is symmetric in an interchange of the
coordinates ${\bf{r}}_{1}$ and ${\bf{r}}_{2}$.

\emph{Triplet $2^{3} S$ State}
% Equation A3
\begin{eqnarray}
\psi ({\bf{r}}_{1}, {\bf{r}}_{2}) &=& Ne^{i m \theta -
{\sqrt{k_\mathrm{eff}}} (r^{2}_{1} + r^{2}_{2})/2} [|{\bf{r}}_{1} -
{\bf{r}}_{2}| + c_{2} |{\bf{r}}_{1} - {\bf{r}}_{2}|^{2} + c_{3}
|{\bf{r}}_{1} - {\bf{r}}_{2}|^{3} + c_{4} |{\bf{r}}_{1} -
{\bf{r}}_{2}|^{4} ] \nonumber \\
N &=& 0.022466 ~~~ \mbox{(Normalization constant)}  \nonumber \\
m &=& 1 ~~~ \mbox{(Angular momentum quantum number)} \nonumber \\
k_\mathrm{eff} &=& 0.072217 ~~~\mbox{(Effective force constant)}
\nonumber \\
c_{2} &=& \frac{1}{3}; ~~ c_{3} = - 0.059108; ~~c_{4} = -0.015884
~~~\mbox{(Coefficients of expansion)} \nonumber \\
E &=& 1.612392 ~(a.u.)^{\star} ~~~ \mbox{(Energy)}.
\end{eqnarray}
Note that since the spin component $\chi (\sigma_{1}, \sigma_{2})$
is symmetric in an interchange of the coordinates $\sigma_{1},
\sigma_{2}$, the spatial component $\psi ({\bf{r}}_{1},
{\bf{r}}_{2})$ is antisymmetric in an interchange of ${\bf{r}}_{1}$
and ${\bf{r}}_{2}$.  That this is the case results from the presence
of the phase factor $e^{i m \theta}$.  When ${\bf{r}}_{1}$ and
${\bf{r}}_{2}$ are interchanged, the magnitude of the relative
vector ${\bf{s}} = {\bf{r}}_{2} - {\bf{r}}_{1}$ does not change, but
its angle $\theta$ (which points from the tip of ${\bf{r}}_{1}$ to
the tip of ${\bf{r}}_{2}$) changes to $\theta + \pi$.  This changes
the sign of the phase factor $e^{i m \theta}$.

\clearpage

% Figure 1
\begin{figure}
\includegraphics[bb=00 00 600 780, width=0.9\textwidth]{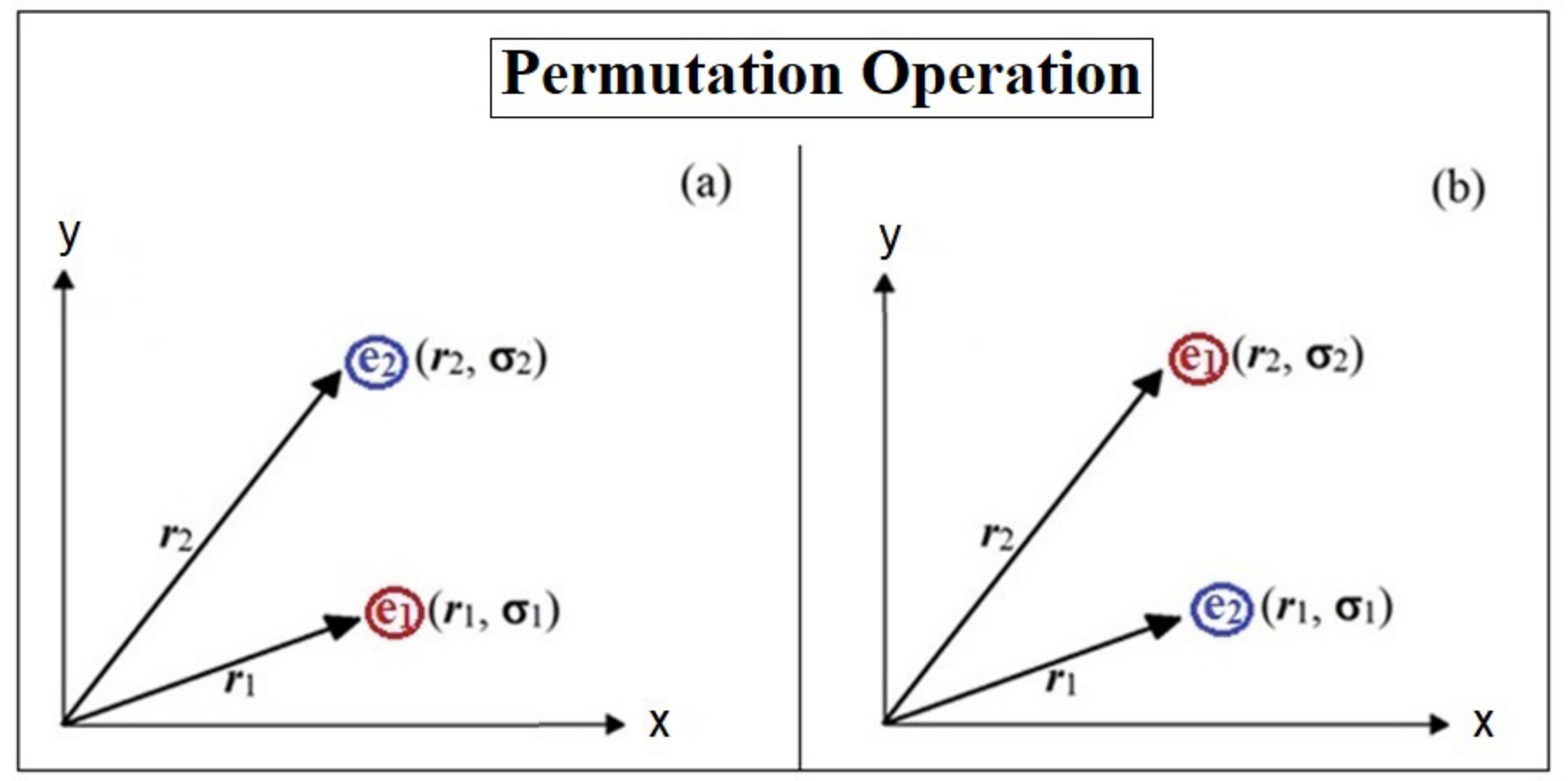}
\caption{Vector description of the permutation operation of the
electronic coordinates. (a) The coordinates before switching are
spatial ${\bf{r}}_{1}$ and spin $\sigma_{1}$ coordinates for
electron 1, and spatial ${\bf{r}}_{2}$ and spin $\sigma_{2}$
coordinates for electron 2; thus $e_{1} ({\bf{r}}_{1}, \sigma_{1})$
and $e_{2} ({\bf{r}}_{2}, \sigma_{2})$. (b)  As a result of
switching both the spatial and spin coordinates for each electron
the new coordinates are spatial ${\bf{r}}_{2}$ and spin $\sigma_{2}$
coordinates for electron 1, and spatial ${\bf{r}}_{1}$ and spin
$\sigma_{1}$ coordinates for electron 2; thus $e_{1} ({\bf{r}}_{2},
\sigma_{2})$ and $e_{2} ({\bf{r}}_{1}, \sigma_{1})$.}\
\\

\end{figure}

% Figure 2
\begin{figure}
\includegraphics[bb=00 00 600 780, width=0.9\textwidth]{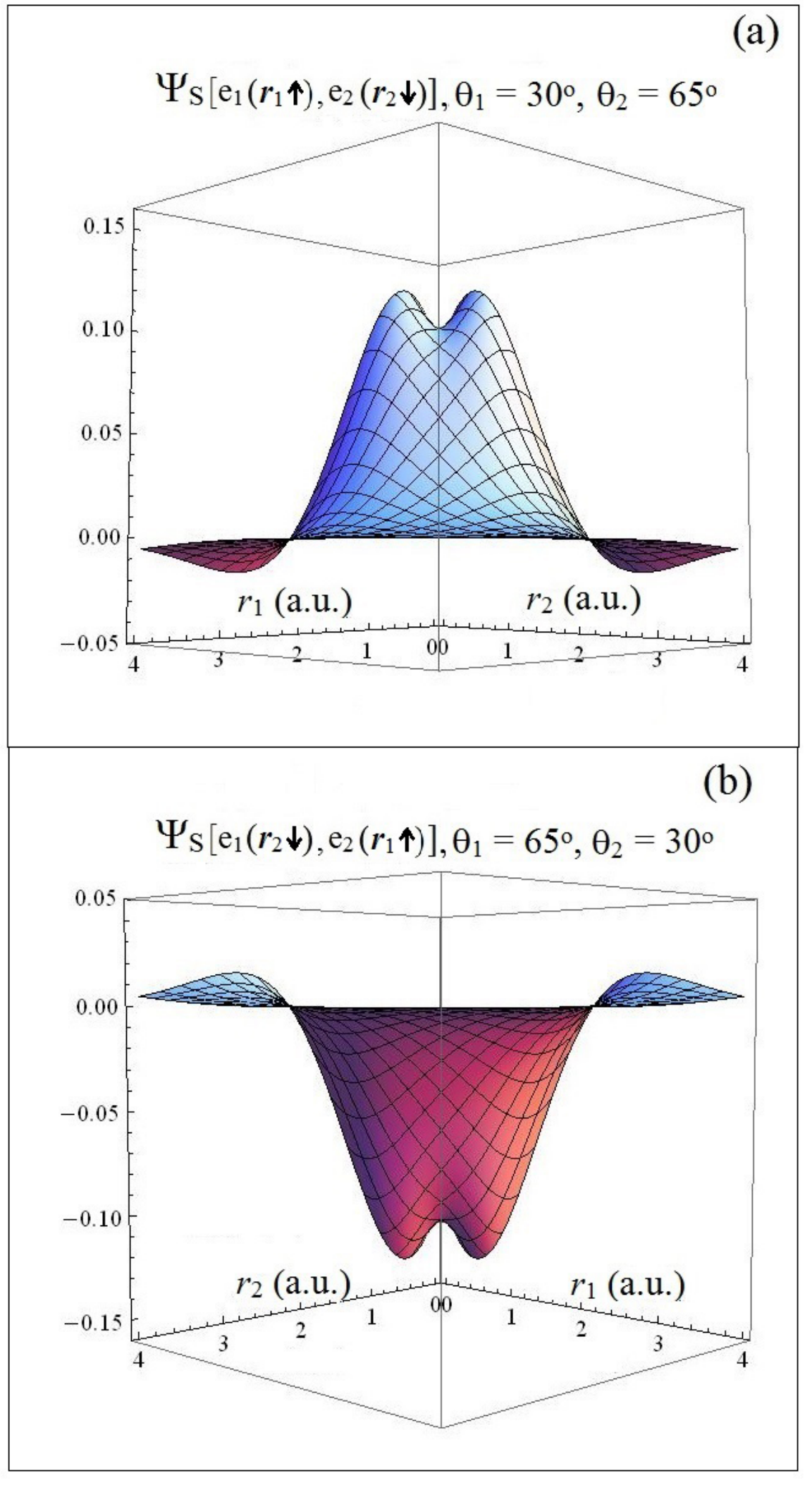}
\caption{Satisfaction of the Pauli principle for the singlet
$2^{1}S$ state wave function of the `artificial atom': (a)
$\Psi_{S}[e_{1}({\bf{r}}_{1} {\boldsymbol{\uparrow}})$,
$e_{2}({\bf{r}}_{2} {\boldsymbol{\downarrow}})](\theta_{1} =
30^{\circ}, \theta_{2} = 65^{\circ})$ ; (b)
$\Psi_{S}[e_{1}({\bf{r}}_{2} {\boldsymbol{\downarrow}})$,
$e_{2}({\bf{r}}_{1} {\boldsymbol{\uparrow}})](\theta_{1} =
65^{\circ}, \theta_{2} =30^{\circ})$. (Note the switch of the
coordinate axes labels in Fig. 2(b).)}\
\\

\end{figure}

% Figure 3
\begin{figure}
\includegraphics[bb=00 00 600 780, width=0.9\textwidth]{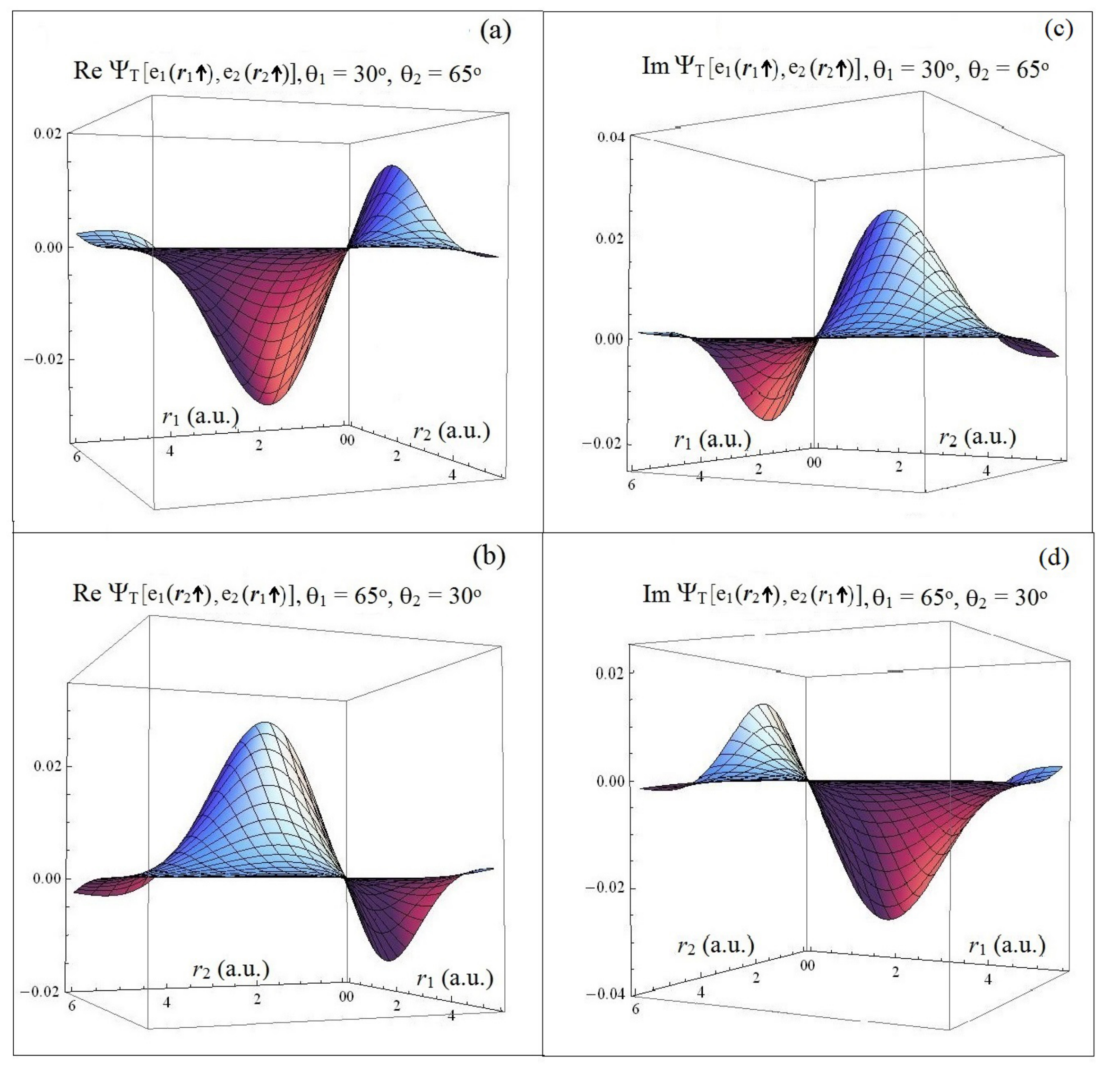}
\caption{Satisfaction of the Pauli principle for the Real and
Imaginary parts of the triplet $2^{3}S$ state wave function for the
`artificial atom': (a) $\Re \Psi_{T} [e_{1}({\bf{r}}_{1}
{\boldsymbol{\uparrow}}), e_{2}({\bf{r}}_{2}
{\boldsymbol{\uparrow}})](\theta_{1} = 30^{\circ}, \theta_{2} =
65^{\circ})$ ; (b) $\Re \Psi_{T} [e_{1}({\bf{r}}_{2}
{\boldsymbol{\uparrow}}), e_{2}({\bf{r}}_{1}
{\boldsymbol{\uparrow}})]$ $(\theta_{1} = 65^{\circ}, \theta_{2}
=30^{\circ})$. (Note the switching of the coordinate axes labels In
Fig. 3(b).)  (c) $\Im \Psi_{T}[e_{1}({\bf{r}}_{1}
{\boldsymbol{\uparrow}}), e_{2}({\bf{r}}_{2}
{\boldsymbol{\uparrow}})]$ $(\theta_{1} = 30^{\circ}, \theta_{2} =
65^{\circ})$; (d) $\Im \Psi_{T}[e_{1} ({\bf{r}}_{2}
{\boldsymbol{\uparrow}}), e_{2}({\bf{r}}_{1}
{\boldsymbol{\uparrow}})]$ $(\theta_{1} = 65^{\circ}, \theta_{2}
=30^{\circ})$. (Note the switching of the coordinate axes labels in
Fig. 3(d).) }\
\\

\end{figure}

% Figure 4
\begin{figure}
\includegraphics[bb=00 00 800 980, width=0.9\textwidth]{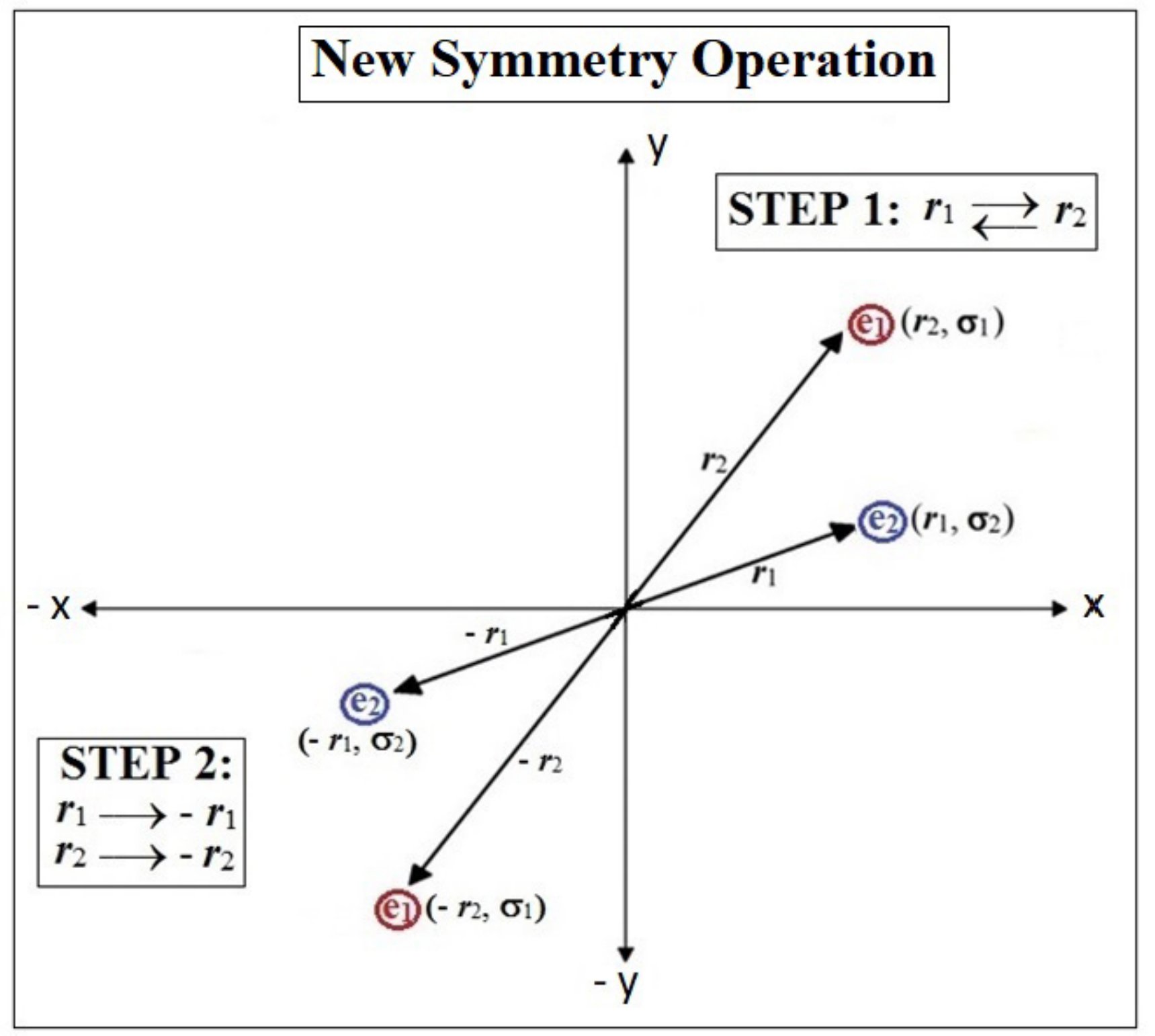}
\caption{Vector description to illustrate the new symmetry
operation. The original electronic coordinates are those described
in Fig. 1 (a). The symmetry operation is a two-step process. STEP 1
(quadrant 1): The spatial coordinates are interchanged while the
spin coordinates remain unchanged which results in spatial
${\bf{r}}_{2}$ and spin $\sigma_{1}$ coordinates for electron 1, and
spatial ${\bf{r}}_{1}$ and spin $\sigma_{2}$ coordinates for
electron 2; thus $e_{1} ({\bf{r}}_{2}, \sigma_{1})$ and $e_{2}
({\bf{r}}_{1}, \sigma_{2})$. STEP 2 (quadrant 3) is an inversion
about the origin. Thus the final result is spatial $-{\bf{r}}_{2}$
and spin $\sigma_{1}$ coordinates for electron 1, and spatial
$-{\bf{r}}_{1}$ and spin $\sigma_{2}$ coordinates for electron 2;
hence $e_{1} (-{\bf{r}}_{2}, \sigma_{1})$ and $e_{2} (-{\bf{r}}_{1},
\sigma_{2})$. }\
\\

\end{figure}

% Figure 5
\begin{figure}
\includegraphics[bb=00 00 600 780, width=0.9\textwidth]{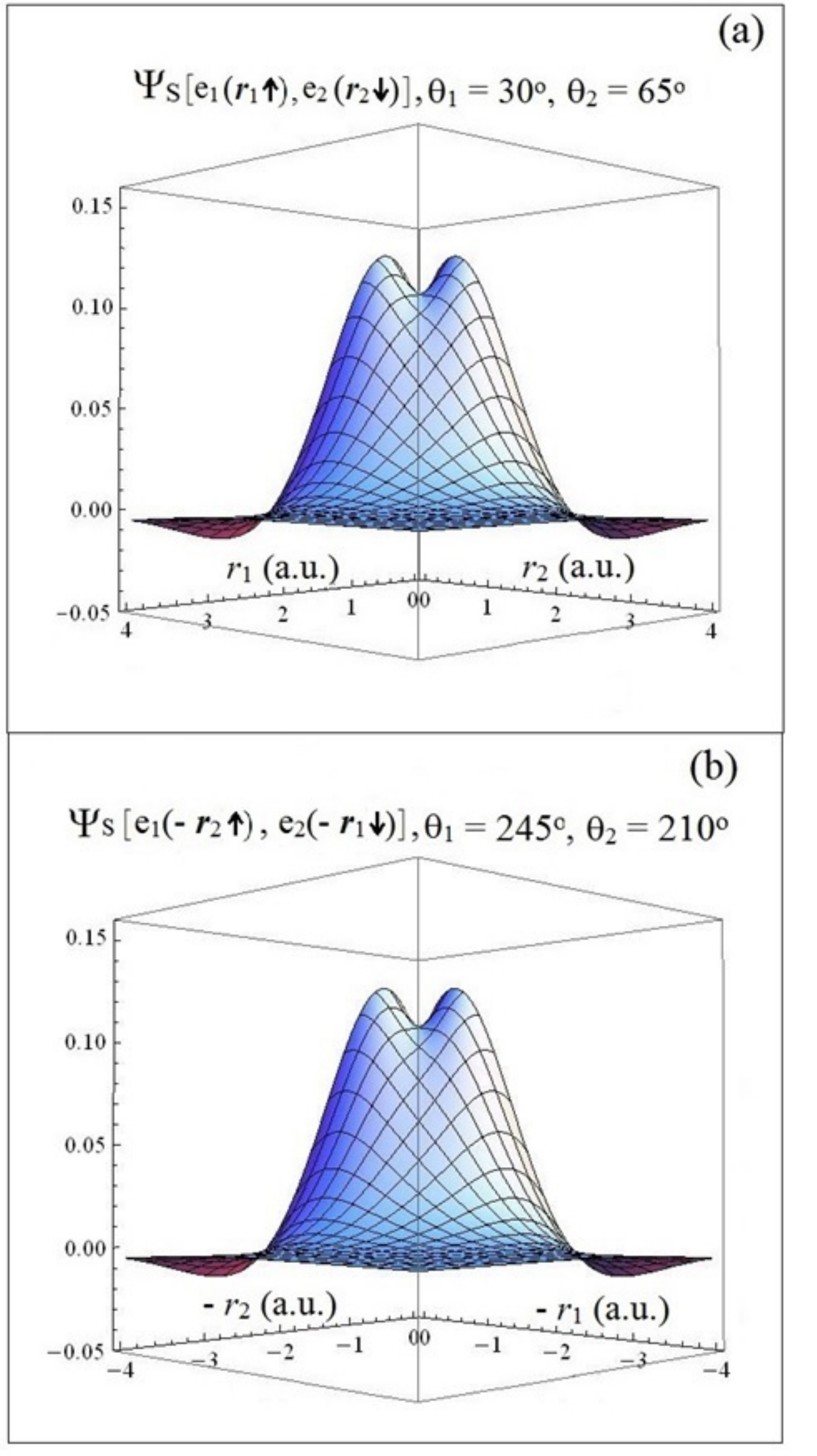}
\caption{Satisfaction of the Wave Function Identity for the singlet
$2^{1}S$ state of the `artificial atom': (a)
$\Psi_{S}[e_{1}({\bf{r}}_{1} {\boldsymbol{\uparrow}}),
e_{2}({\bf{r}}_{2} {\boldsymbol{\downarrow}})]$ $(\theta_{1} =
30^{\circ}, \theta_{2} = 65^{\circ})$ and (b)
$\Psi_{S}[e_{1}(-{\bf{r}}_{2} {\boldsymbol{\uparrow}}),
e_{2}(-{\bf{r}}_{1} {\boldsymbol{\downarrow}})]$ $(\theta_{1} =
245^{\circ}, \theta_{2} = 210^{\circ})$.}\
\\

\end{figure}

% Figure 6
\begin{figure}
\includegraphics[bb=00 00 600 780, width=0.9\textwidth]{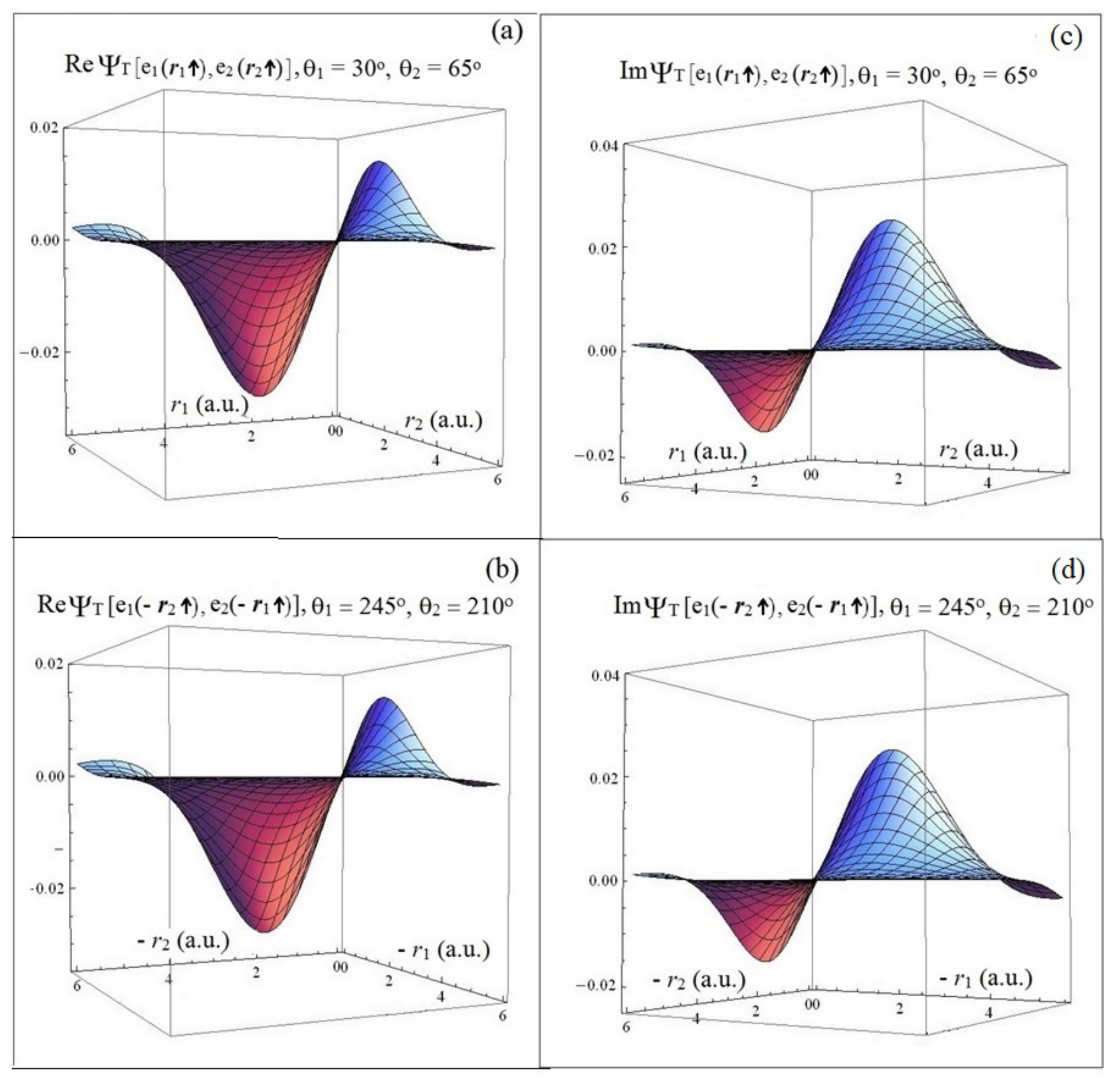}
\caption{Satisfaction of the Wave Function Identity for the Real and
Imaginary parts of the triplet $2^{3}S$ state of the `artificial
atom': (a) $\Re \Psi_{T}[e_{1}({\bf{r}}_{1}
{\boldsymbol{\uparrow}}), e_{2}({\bf{r}}_{2}
{\boldsymbol{\uparrow}})]$ $(\theta_{1} = 30^{\circ}, \theta_{2} =
65^{\circ})$ and (b) $\Re \Psi_{T} [e_{1} (-{\bf{r}}_{2}
{\boldsymbol{\uparrow}}), e_{2}(-{\bf{r}}_{1}
{\boldsymbol{\uparrow}})]$ $(\theta_{1} = 245^{\circ}, \theta_{2} =
210^{\circ})$ and (c)  $\Im \Psi_{T}[e_{1}({\bf{r}}_{1}
{\boldsymbol{\uparrow}}), e_{2}({\bf{r}}_{2}
{\boldsymbol{\uparrow}})]$ $(\theta_{1} = 30^{\circ}, \theta_{2} =
65^{\circ})$ and (d) $\Im \Psi_{T}[e_{1}(-{\bf{r}}_{2}
{\boldsymbol{\uparrow}}), e_{2}(-{\bf{r}}_{1}
{\boldsymbol{\uparrow}})]$ $(\theta_{1} = 245^{\circ}, \theta_{2} =
210^{\circ})$. }\
\\

\end{figure}

% Figure 7
\begin{figure}
\includegraphics[bb=00 00 600 780, width=0.9\textwidth]{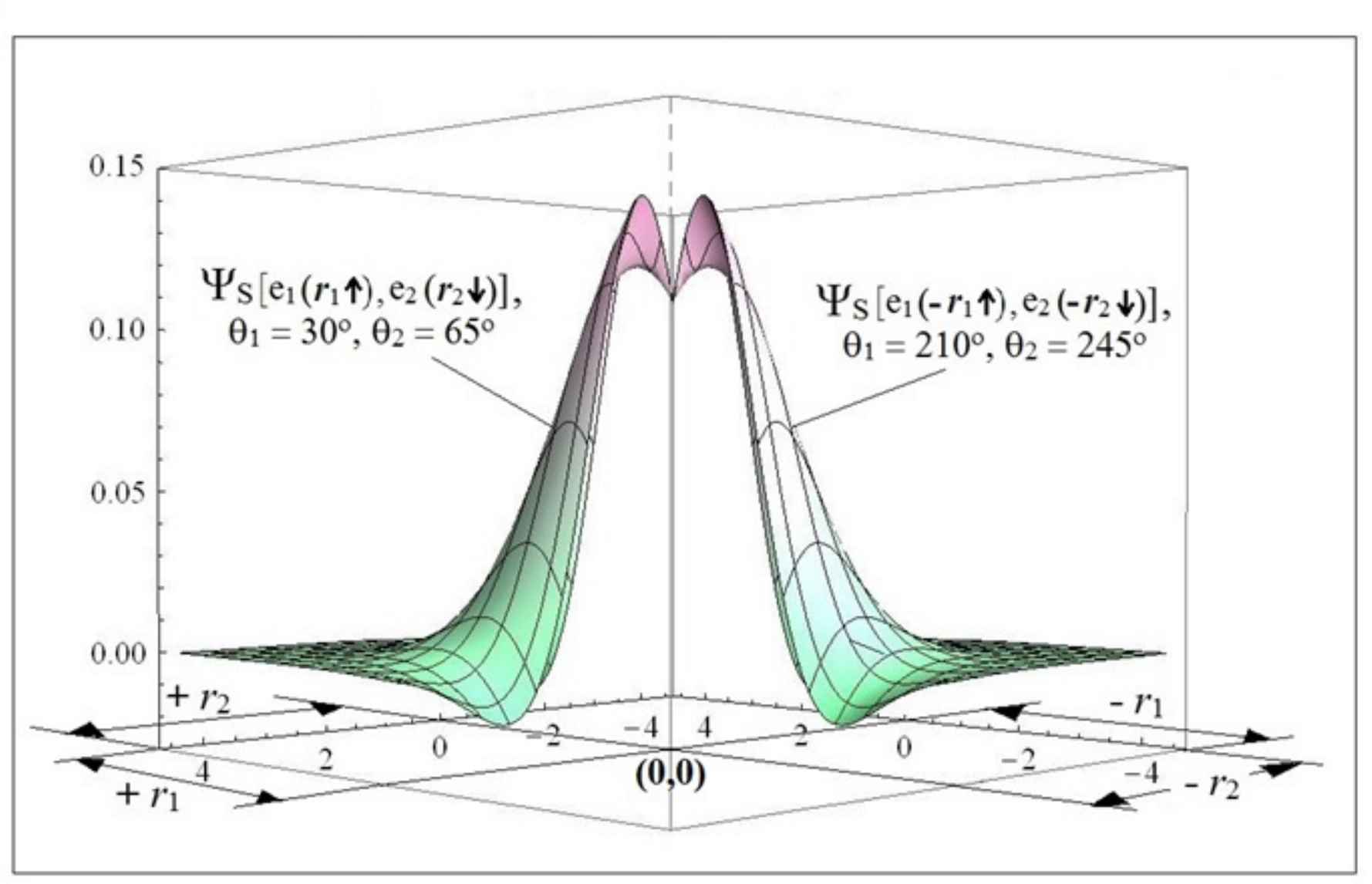}
\caption{Even Parity of the singlet $2^{1}S$ state wave function of
the `artificial atom'.  The functions $\Psi_{S}[e_{1}({\bf{r}}_{1}
{\boldsymbol{\uparrow}}), e_{2}({\bf{r}}_{2}
{\boldsymbol{\downarrow}})]$ $(\theta_{1} = 30^{\circ}, \theta_{2} =
65^{\circ})$ and $\Psi_{S} [e_{1}(-{\bf{r}}_{1}
{\boldsymbol{\uparrow}}), e_{2}(-{\bf{r}}_{2}
{\boldsymbol{\downarrow}})]$ $(\theta_{1} = 210^{\circ}, \theta_{2}
= 245^{\circ})$ are plotted.}\
\\

\end{figure}

% Figure 8
\begin{figure}
\includegraphics[bb=00 00 600 780, width=0.9\textwidth]{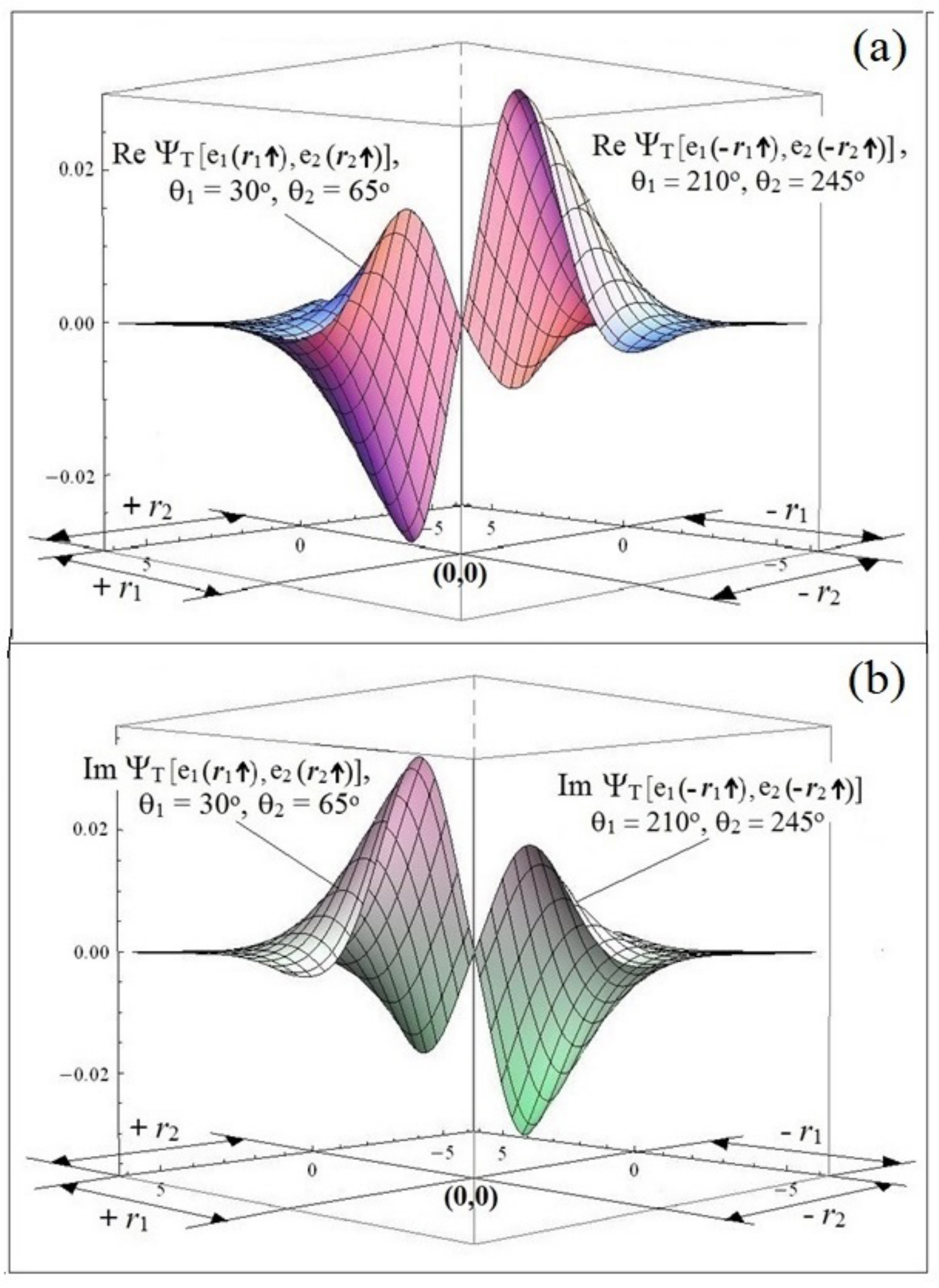}
\caption{Odd Parity of the Real and Imaginary parts of the triplet
$2^{3}S$ wave function for the `artificial atom'. The following
functions are plotted  (a)$\Re \Psi_{T}[e_{1}({\bf{r}}_{1}
{\boldsymbol{\uparrow}}), e_{2}({\bf{r}}_{2}
{\boldsymbol{\uparrow}})](\theta_{1} = 30^{\circ}, \theta_{2} =
65^{\circ})$ and $\Re \Psi_{T}[e_{1}(-{\bf{r}}_{1}
{\boldsymbol{\uparrow}}), e_{2}(-{\bf{r}}_{2}
{\boldsymbol{\uparrow}})](\theta_{1} = 210^{\circ}, \theta_{2} =
245^{\circ})$; (b)  $\Im \Psi_{T} [e_{1}({\bf{r}}_{1}
{\boldsymbol{\uparrow}}), e_{2}({\bf{r}}_{2}
{\boldsymbol{\uparrow}})](\theta_{1} = 30^{\circ}, \theta_{2} =
65^{\circ})$ and $\Im \Psi_{T}[e_{1}(-{\bf{r}}_{1}
{\boldsymbol{\uparrow}}), e_{2}(-{\bf{r}}_{2}
{\boldsymbol{\uparrow}})](\theta_{1} = 210^{\circ}, \theta_{2} =
245^{\circ})$. }\
\\

\end{figure}

% Figure 9
\begin{figure}
\includegraphics[bb=00 00 600 780, width=0.9\textwidth]{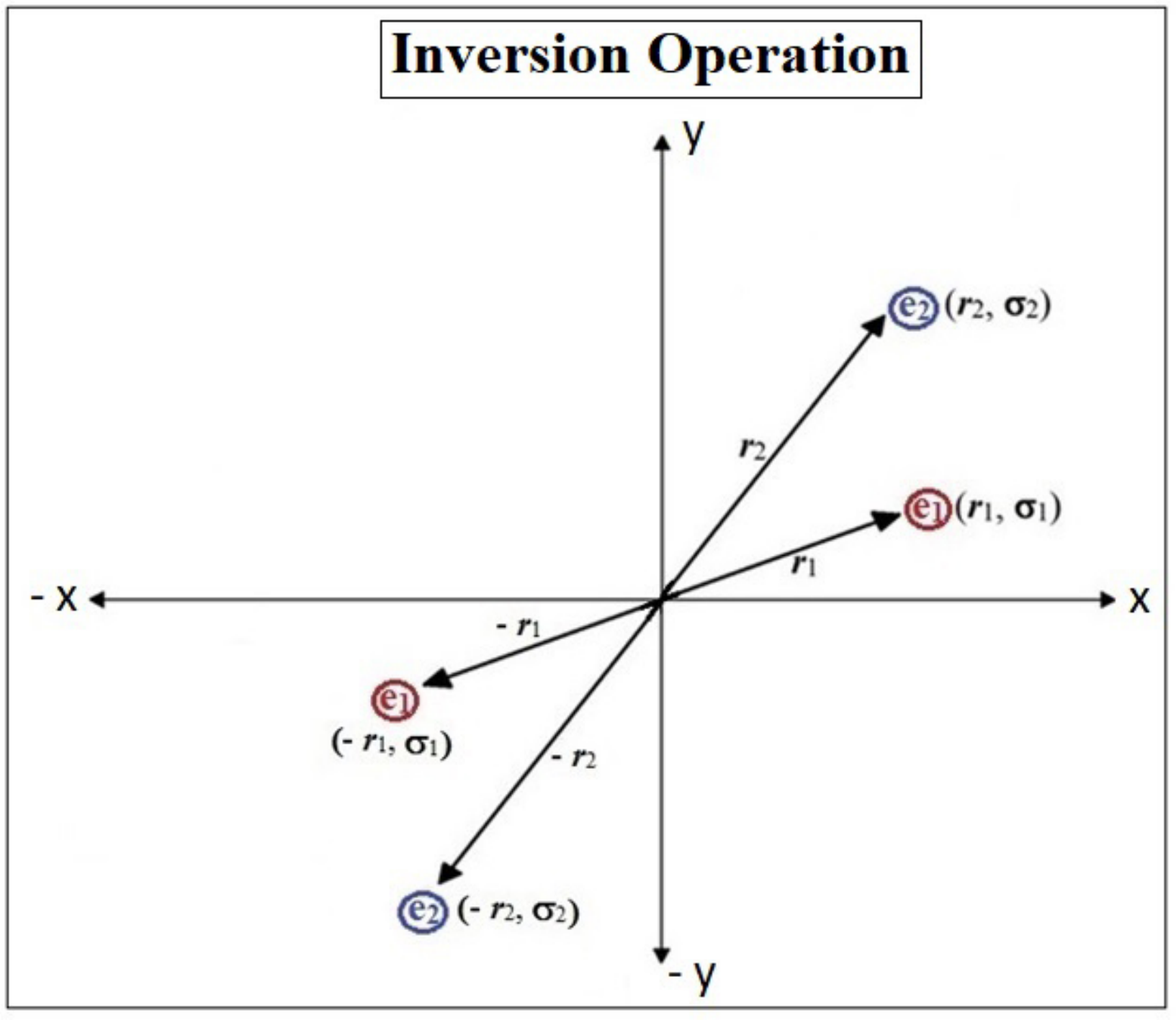}
\caption{Vector description of the inversion operation.  The figure
is valid for both the singlet and triplet states.  Quadrant 1 is the
same as Fig. 1 (a).  Quadrant 3: Electron 1 with spatial
$-{\bf{r}}_{1}$ and spin $\sigma_{1}$ coordinates: $e_{1}
(-{\bf{r}}_{1}, \sigma_{1})$; Electron 2 with spatial
$-{\bf{r}}_{2}$ and spin $\sigma_{2}$ coordinates:
$e_{2}(-{\bf{r}}_{2}, \sigma_{2})$.}\
\\

\end{figure}

% Figure 10
\begin{figure}
\includegraphics[bb=00 00 600 780, width=0.9\textwidth]{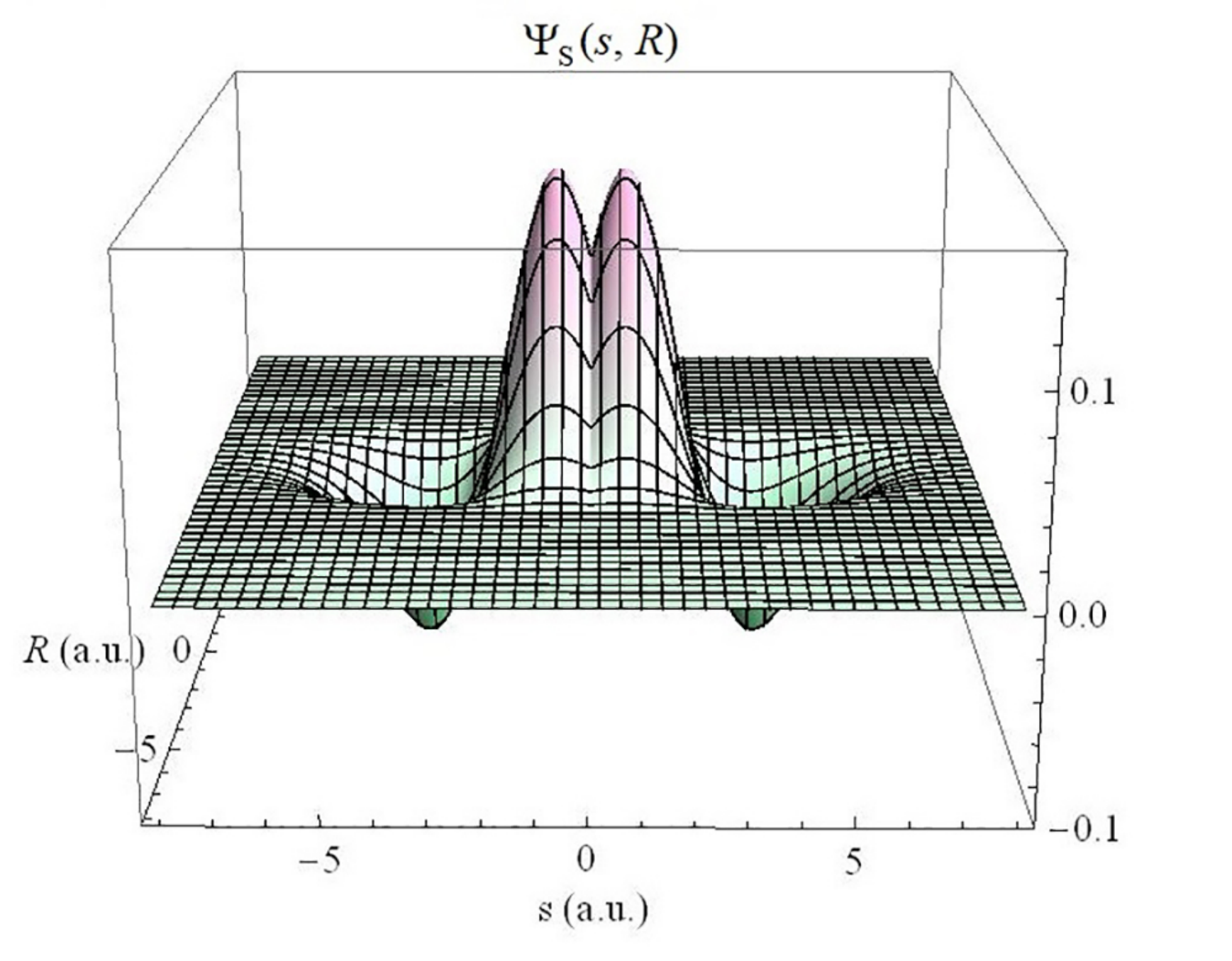}
\caption{The singlet $2^{1}S$ state wave function of the `artificial
atom' plotted as a function of the center of mass ${\bf{R}}$ and
relative ${\bf{s}}$ coordinates.}\
\\

\end{figure}

% Figure 11
\begin{figure}
\includegraphics[bb=00 00 600 780, width=0.9\textwidth]{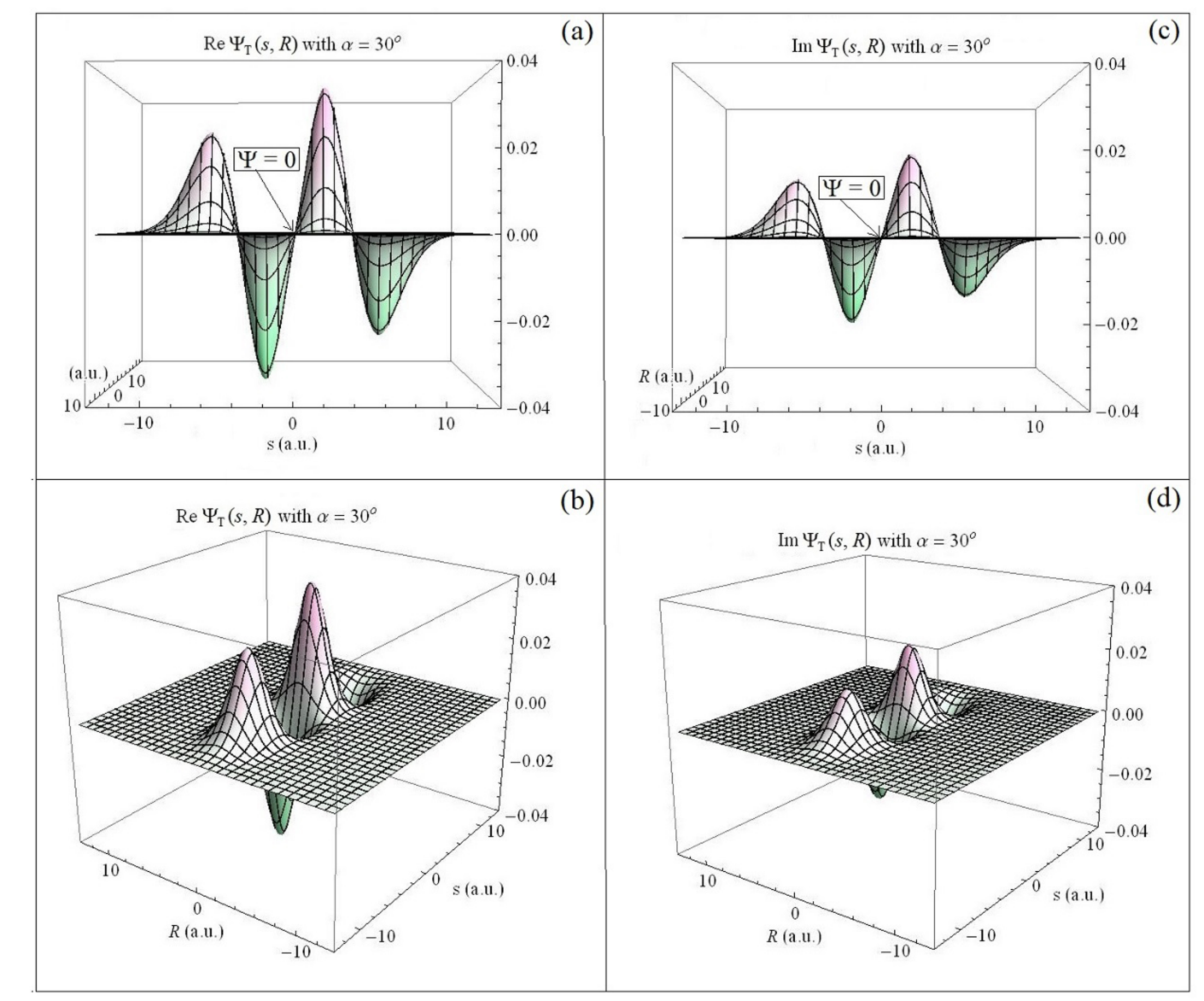}
\caption{Two views of the Real (a), (b) and Imaginary (c), (d) parts
of the triplet $2^{3}S$ wave function for the `artificial atom'
plotted as a function of the center of mass ${\bf{R}}$ and relative
${\bf{s}}$ coordinates for $\alpha = 30^{\circ}$, where $\alpha$ is
the angle of the relative coordinate ${\bf{s}}$.}\
\\

\end{figure}

\end{document}